\RequirePackage{fix-cm}
\documentclass[journal]{IEEEtran}
\usepackage{epsfig,graphics,graphicx,subfig,amssymb,amstext,amsmath,algorithm,algorithmic,wrapfig,multirow}
\usepackage{array}
\usepackage{cite}
\usepackage{url}
\usepackage{times}
\usepackage{lipsum}
\usepackage{float}
\usepackage{placeins}
\usepackage{textcomp}
\usepackage[font=small]{caption}
\usepackage{hyperref}
\usepackage{flushend}
\usepackage{xcolor}
\usepackage{tabularx}
\usepackage{bm}
\usepackage{enumerate}
\usepackage{tikz}
\usepackage{pgfplots}
\usepackage{enumerate}
\usepackage{epstopdf}
\usetikzlibrary{shapes,arrows}
\usepackage[utf8]{inputenc}
\usepackage{mathtools}
\usepackage[euler]{textgreek}

\usepackage{color}

\def\beq{\begin{equation}}
\def \eeq{\end{equation}}
\def\beqa{\begin{eqnarray}}
\def\eeqa{\end{eqnarray}}
\def\beqan{\begin{eqnarray*}}
\def\eeqan{\end{eqnarray*}}

\def\C{{\mathbb{C}}}
\def\argmin{\mathop{\mathrm{arg\,min}}}

\def\Exp{\mathbb{E}}

\def\tm1{t\! - \! 1}
\def\tp1{t\! + \! 1}

\def\nbf{\mathbf{n}}

\def\ubf{\mathbf{u}}

\def\vbf{\mathbf{v}}

\def\xbf{\mathbf{x}}

\def\ybf{\mathbf{y}}

\def\Hbf{\mathbf{H}}

\def\Ibf{\mathbf{I}}

\def\alphahat{\widehat{\alpha}}
\def\alphabf{{\boldsymbol \alpha}}

\def\ellhat{\widehat{\ell}}

\def\sigmahat{\widehat{\sigma}}
\def\sigmabf{{\boldsymbol \sigma}}

\renewcommand{\footnoterule}{
 \kern -3pt
 \hrule width \columnwidth height 0.5pt
 \kern 3pt
}

\begin{document}

\title{Energy and Latency of Beamforming Architectures for Initial Access in mmWave Wireless Networks
}

\author{
        C. Nicolas Barati,
        Sourjya Dutta,
        Sundeep Rangan
        and Ashutosh Sabharwal

        \thanks{
            C. Nicolas Barati (email: nicobarati@rice.edu) and
            Ashutosh Sabharwal (email: ashu@rice.edu)
            are with the Department of Electrical and Computer Engineering,
            Rice Univercity, Houston, TX.
            Sourjya Dutta (email: sdutta@nyu.edu),
            Sundeep Rangan (email: srangan@nyu.edu),
            are with the NYU WIRELESS, Tandon School of Engineering,
            New York University, Brooklyn, NY.
         }
    }

\maketitle

\begin{abstract}
Future millimeter-wave (mmWave) systems, 5{G} cellular or WiFi,
must rely on highly directional links to
overcome severe pathloss in these frequency bands.
Establishing such links requires the mutual discovery
of the transmitter and the receiver 
potentially leading to a large latency and high energy consumption.
In this work, we show that both the discovery latency and energy
consumption can be significantly reduced by using fully digital front-ends.
In fact, we establish that by reducing the resolution of the fully-digital front-ends we can achieve lower energy consumption compared to both analog and high-resolution digital beamformers.
Since beamforming through analog front-ends allows sampling in only
one direction at a time, the mobile device is ``on''
for a longer time compared to a
digital beamformer which can get spatial samples from all directions in one shot.
We show that the energy consumed by the analog front-end can be four to six times more than that of the digital front-ends,
depending on the size of the employed antenna arrays.
We recognize, however, that using fully digital beamforming post beam discovery, i.e., for data transmission,
is not viable from a power consumption standpoint.
To address this issue, we propose the use of digital beamformers
with low-resolution analog to digital converters ($4$ bits).
This reduction in resolution brings the power consumption to the same level as analog beamforming for data transmissions while benefiting from the spatial multiplexing capabilities of fully digital beamforming, thus reducing initial discovery latency and improving energy efficiency. 

 
\end{abstract}


\section{Introduction}
\label{intro}


Almost all of the current wireless communication takes place in a relatively
small region of the electromagnetic spectrum below $6 {\rm ~GHz}$.
This region has been allocated by government agencies around the world for commercial, civilian, military, public safety and experimental use.
However, the proliferation of devices and services that use or depend on wireless technologies has caused an ever-increasing discrepancy between the demand and the available bandwidth, or degrees of freedom (DoF). 
This discrepancy termed {\it spectrum crunch}, if not addressed, will lead to lower data rates and reduced quality of service.
Spectrum crunch will become even more acute
when Internet of Things \cite{qc_iot_whitepaper} and Device to Device communication traffic are
added to the already overloaded networks.

Increasing the DoFs is the only option for the
next generation (5{G}) wireless systems. 
The use of mmWave enables an increase in DoFs by adding more bandwidth due to the availability of large (in the order of a few GHz) unlicensed spectrum between $30$ and $300~{\rm GHz}$.
However, as explained by Friis' law \cite{Rappaport:02}, signals transmitted
in mmWaves have high isotropic pathloss, i.e., they decay at a much higher rate with the traveled distance.
This leads to a reduced communication range compared to sub-$6~$GHz systems. 
Furthermore, mmWaves exhibit characteristics resembling the visible light. For example, they have high penetration loss through most material
and hence are easily blocked by the surrounding objects.
MmWave systems can overcome these shortcomings by employing
beamforming (BF). That is, to use arrays of multiple antenna elements to extend the communication range and avoid obstacles in the environment
by directing the signal energy in an intended direction.

However, the reliance of mmWave communication on directional links
through beamforming poses new challenges that do not
exist in wireless systems over the microwave bands. The transmitter (Tx) and the receiver (Rx) must first discover each other directions before they can start the data communication.
This process, known in cellular systems as \emph{initial access}, is generally performed omni-directionally (or using very wide beams) in the sub-$6~$GHz bands.
But, due to the high path loss, if mmWave systems were to follow the same paradigm, the range of
mutual discovery would be much smaller than the range where directional
high-rate communication would be possible.
Therefore, mutual discovery must be performed in a directional manner.

The directional discovery phase can last for a long time when the base-station and the user employ arrays with many antenna elements forming narrow beams. 
While searching for the base-station, the battery-limited user
is always ``on'' burning energy.
We show here that this energy consumption can be reduced significantly
by employing a low-resolution fully digital front-end on the user side.
The reason is, beamforming through a digital front-end reduces the discovery
latency (or delay) by an order of magnitude compared to an analog front-end.
Hence, the user is ``on'' for a shorter span of time leading to considerable energy savings.

While the focus of this paper is directional discovery in initial access,
directional discovery is expected to be triggered also in other phases of
mmWave communication.
For example, in recovery from link failures, which will be frequent due to the sensitivity of the mmWave links to obstacles and changes in the environment.
Handovers to a new base-station will also be frequent since
mmWave cells will be smaller in size compared to the current ones and
a mobile user may go through more cells for the same distance traveled.  
More importantly, since the battery-dependent devices operate at such high frequencies and bandwidth, more aggressive use of sleep/idle mode (discontinuous reception) will be critical from an energy consumption standpoint.
For each of these operations, the user device must often (re-) discover the direction to the connected and neighboring base-station(s).
Therefore, it is extremely crucial that directional discovery and beam alignment is fast and energy-efficient.

\subsection{Contributions}
\label{sec:contr}

In this paper, we look into the problem of latency and energy consumption
in directional discovery in mmWave systems during initial access. 
Our focus is on 5G cellular systems where the issue of communication range is more important and challenging than short-range mmWave communication, e.g., 802.11ad WiFi.
There are two key take-away points in our work:

\begin{itemize}
  \item[1.] Digital beamforming results in both low latency and
  low energy consumption during initial discovery
  compared to analog beamforming.
  \item[2.] Employing low resolution analog to digital converters (ADCs) in fully digital front-ends can achieve low latency and even lower power consumption for both control signalling and data transmissions.
\end{itemize}

Mutual directional discovery requires the transmitter and the receiver
search in their surrounding angular space at a granularity inversely
proportional to the size of their antenna arrays: larger antenna arrays
result in narrower beams potentially leading to higher discovery delays.
In a cellular environment, this latency
will also affect user handovers between different base-stations,
alternative link discovery in case of blockage and
the ``idle mode'' to ``on mode'' circles.

We show that due to the large latency in directional discovery incurred by analog beamforming, its energy consumption is greater not only than low-resolution digital but also than high-resolution digital. 
This difference in energy consumption increases with the size of the
antenna arrays. When a $4$-by-$4$ antenna array is used, analog beamforming
burns as much as six times more energy than digital.
This is due to the fact that analog beamforming needs more
time to sample an angular domain that increases in size
with the number of antennas.

Leveraging our previous work
\cite{barati2015directional, barati2016initial}, we establish a
relationship between the beamforming
architecture, and the mutual discovery delay within the context of 3GPP
initial access for mmWave networks \cite{3GPP_NRphyRRC}.
Specifically, we show that between analog and digital beamforming,
digital outperforms analog by a large margin -- in the order
of $300$ to $900 {\rm ms}$.
Similar to \cite{dutta17Asilomar} and \cite{dutta2019twc}, we detail and compare the power consumption of four beamforming architectures, namely,
analog, digital, low-resolution digital and hybrid, by assessing the
components and devices they are comprised of.
It is known that by reducing the resolution of
analog to digital converters (ADC)
we can significantly reduce the power consumption of a fully
digital beamforming circuit.However, we show that while this reduction
comes at a penalty of less than $1 ~ {\rm dB}$
SNR in the low-SNR regime, the discovery delay is kept at the same low levels as digital beamforming with high resolution -- $20$ to $80~{\rm ms}$. Thus, low resolution fully digital beamformers outperform analog as they can be power efficient during both data transmissions as well as signaling control messages.

  

  

Interestingly, in most studies related to mmWaves systems analog
beamforming has been
preferred over digital due to its low device power consumption.
However, as we show, when discovery delay in initial access is taken into
account, analog beamforming can burn multiple times more energy than any
other alternative. 


\subsection{Related Work}
\label{sec:relwo}



Due to the reliance on highly directional links at mmWave frequencies, efficient beam management is key to establishing and maintaining a reliable link.
A critical component of beam management is the beam discovery procedure for initial access.
Current technical literature assumes an analog or hybrid front-end which limits the number of usable spatial streams. For instance, \cite{dedonno_2017, palacios_2016} present a heuristic method to design beam patterns for initial beam discovery for hybrid beamformers.
Raghavan et al. in \cite{raghavan2016} proposes to vary the beam widths depending on the user link quality. They show that users with good link (high SNR) can be detected with wide beams leading to a decrease in detection delays.
In another direction of research, \cite{zhang_code2017,
jaesim_code2019, Hussain_code2017} aim at designing
optimal beam code-books, i.e., sets of directions,
for connection establishment.
Another important area of research has been the use of \emph{side information}, for instance location information, channel quality measurements at microwave bands, etc., has been studied in \cite{capone_cont2015, filipini_cont2018,capone_obst2015, marcano_macrocell2016, devoti_cont2018, abbas_context2016, xiu_outband2019, garcia_inband2018, park_locbase2017}.
Moreover, the work in \cite{sim_mlv2v2018} propose the use of online machine learning algorithms for beam detection for vehicular communications at mmWave.
The works in \cite{zhang_game2016, marandi_game2019, rave_game2019} use Game theoretic methods whereas in \cite{guo_gen2017, souto_gen2019}
the authors use genetic algorithms for initial beam discovery.

In \cite{alkhateeb_init2017} the authors present a theoretical analysis of the trade-off between spending resources for accurate beam alignment on one hand and using them for actual data communication on the other. More interestingly, the analysis in \cite{alkhateeb_init2017} shows that with large coherence block lengths, exhaustive beam search outperforms hierarchical search. This understanding is reflected in the current 3GPP specifications \cite{3GPP_NRphyMod, 3GPP_NRphyRRC, 3GPP_NRphyProcCont, 3GPP_NRphyPhyMeas} on initial access where initial beam discovery and alignment is achieved through exhaustive search. We will discuss the 3GPP New Radio beam search procedure briefly in Sec. \ref{sec:sigsysmodgpp_par}. A detailed overview of the beam management procedure for 5G systems can be found in \cite{giordani2018}.

Critical to beam discovery and beam alignment is the efficient signaling of pilot or synchronization signals and channel estimation.
To this end, the work in \cite{eliasi_lowTWC},
leverages the sparsity of the mmWave channel. The authors use a compressed sensing framework for
estimating the number of measurements necessary for estimating the channel covariance matrix for beam/angle detection with high confidence. Similarly, in \cite{cairecs2018}
compressed sensing is used for fast angle of arrival/departure and
second-order channel statistics estimation. In \cite{yan_cs2016} a compressed sensing-based
algorithm robust to frequency offsets and phase noise is presented. The sparsity of the mmWave channel is exploited
in \cite{booth_bandit2019} where a novel algorithm based on multiple-armed bandit beam selection for both initial beam
alignment and beam tracking is proposed.
In \cite{hashemi_2018_cont},
Hashemi \textit{et. al.} exploit the channel correlation to reduce the searching space and subsequently the delay of beam discovery.

To better understand the interplay between the hardware (and power) constraints at mmWave and the beam discovery delay, in our previous work \cite{barati2015directional} \cite{barati2016initial}, we presented a comparative analysis of analog and digital beamforming in terms of synchronization signal detectability and delay.
We show that digital beamforming, even with low-resolution quantizers, performs dramatically better compared to analog or hybrid.
Furthermore, our recent work, \cite{dutta17Asilomar} and \cite{dutta2019twc}, presents a thorough study of various beamforming architectures at both the transmitter and the receiver sides. There, we show that employing
losses in system rate under practical mmWave cellular assumptions. Based on this, we argue that as low-resolution fully digital front-ends can have \emph{low control delays} while having the same power efficiency as analog and hybrid beamformers for the \emph{data plane}.
To the best of our knowledge, this is the first
work where energy consumption during initial beam discovery has been studied. In this work, we show that a low resolution fully digital beamformer
is energy efficient in both control plane in general, initial access in particular,
and data plane. 

The rest of the paper is organized as follows.
In Sec. \ref{sec:beamdis} we give an overview of the beam
discovery problem and present the 3GPP New Radio (NR)
discovery procedure. In Sec. \ref{sec:sigsysmod} we
present the system model. We derive a correlation-based
detector for beam discovery, present four mmWave beamformers and comment on the power consumption
of each one of them. In Sec. \ref{sec:evsim}, we
evaluate through simulations the performance of the
analog and high/low-resolution beamformers in terms of
discovery delay and energy consumption.

\paragraph{Notation:} We use bold-face small letters $(\mathbf{h})$ to denote vectors
and bold-face capital letters $(\mathbf{H})$ for matrices.
Conjugate and conjugate transpose are denoted with $\mathbf{h}^*$ and the 
$\ell$-2 norm with $\| \mathbf{h} \|$ and $\| \mathbf{H} \|$ for vectors and matrices, respectively.

\section{Beam discovery through beamsweep}
\label{sec:beamdis}

\begin{figure*}
\centering
\subfloat [Sectorized Search] {
	\includegraphics[width=0.4\textwidth]{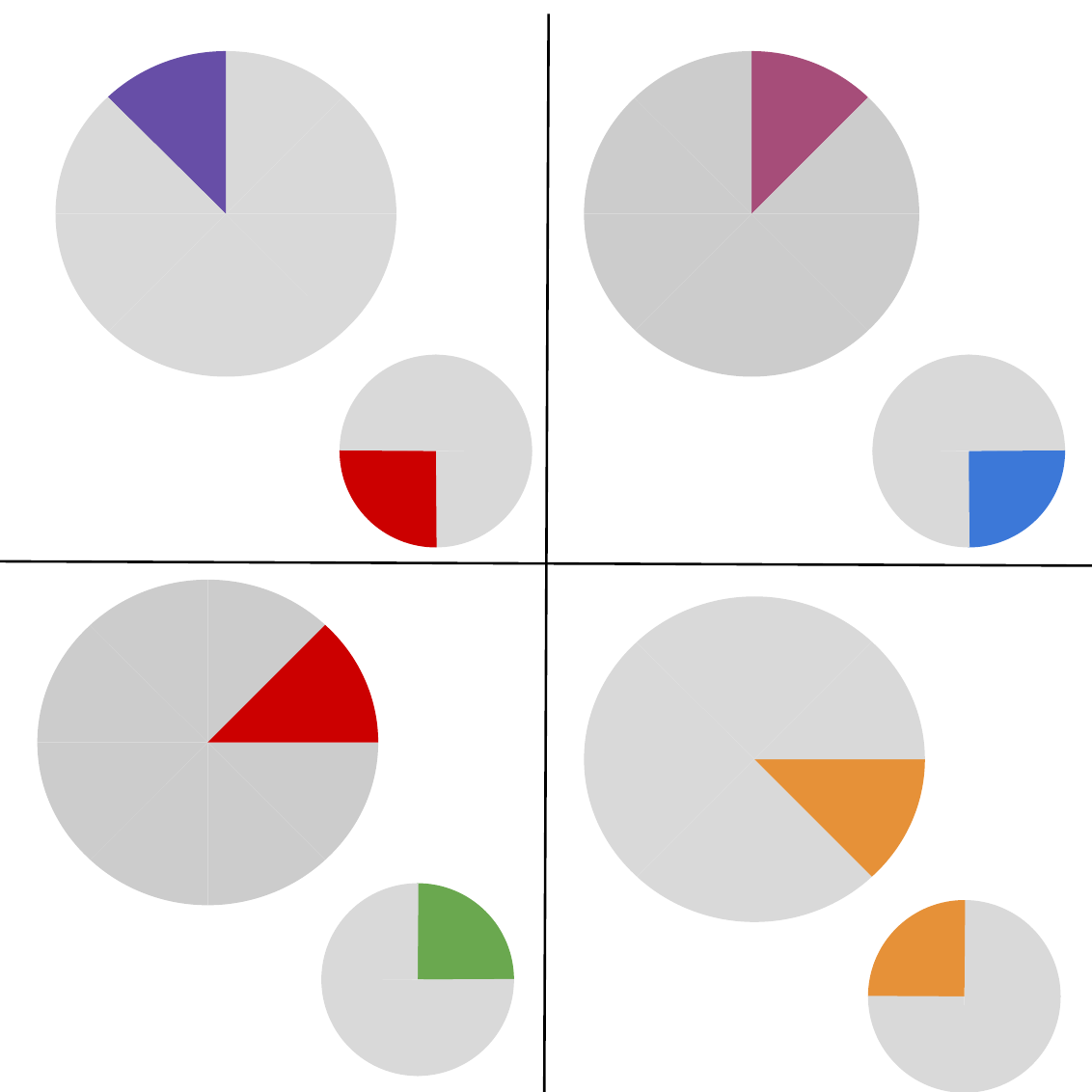} \label{fig:bspace_sec}
} \qquad
\subfloat [Hierarchical Search] {
	\includegraphics[width=0.4\textwidth]{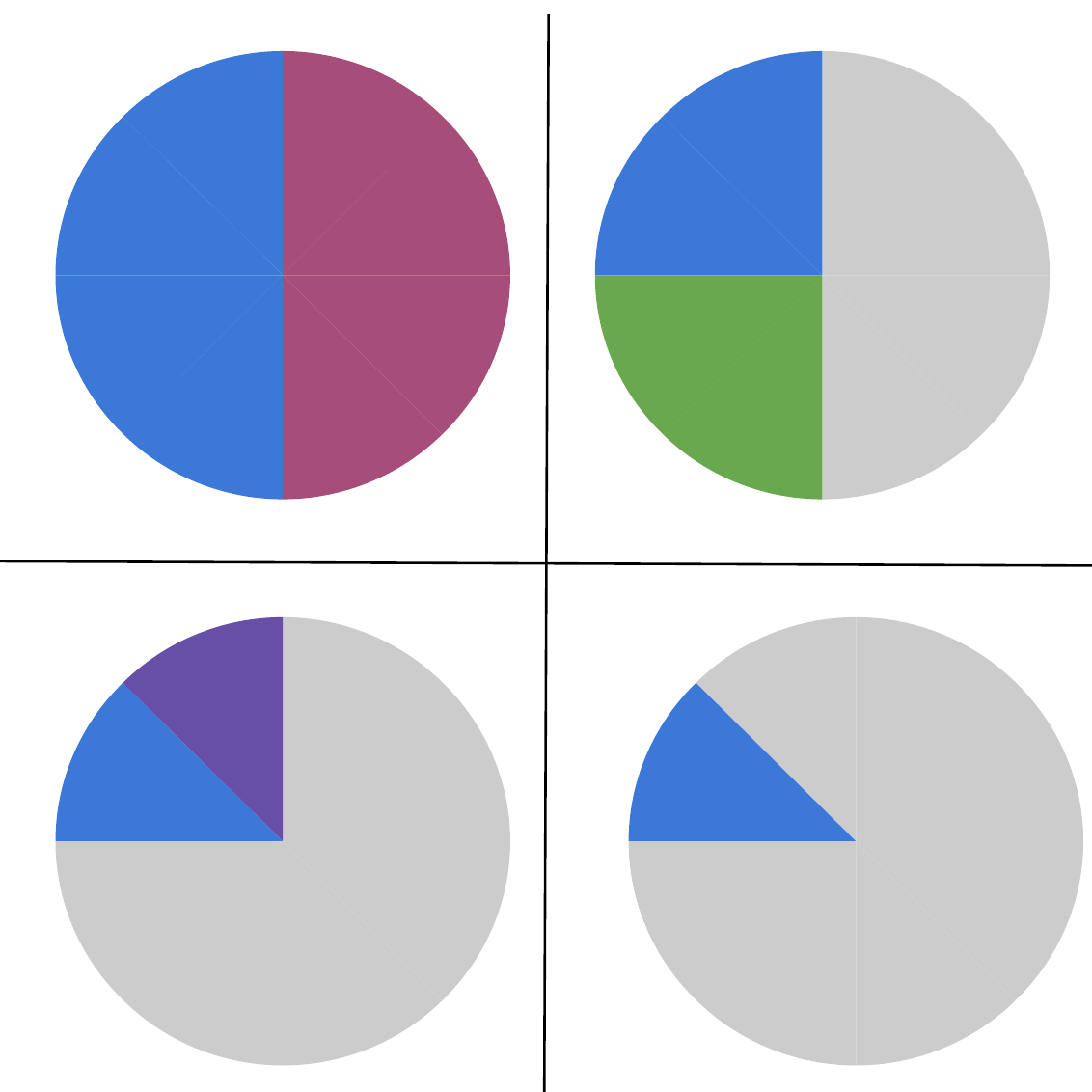} \label{fig:bspace_hier}
}
\caption{Beam discovery in mmWave bands. Panels read from left to right and top to bottom. Fig. (\ref{fig:bspace_sec}) Search over narrow sectors. the receiver, larger shape, and the transmitter, smaller shape, sequentially visit each sector combination until mutual discovery. Fig. (\ref{fig:bspace_hier}) Hierarchical search. the receiver gradually narrows its beams in a series of steps.}
\label{fig:bspace}
\end{figure*}

Consider a transmitter-receiver pair operating in mmWave bands.
Since they communicate through directional links, 
they both must \emph{discover} each other in the spatial domain before the data communication can begin.
An intuitive approach to this problem is to divide the spatial domain around the transmitter (or receiver)
into multiple non-overlapping sectors, as presented in Fig. \ref{fig:bspace_sec}.
Each of these sectors corresponds to a transmitter or receiver beam. All possible combinations of the pair of transmitter-receiver beams is termed as the \emph{beamspace}. 
By increasing the number of antennas makes the beams narrower which increases the beamforming gain but also makes the size of the beamspace larger. This poses a fundamental problem for mmWave systems. On one hand, narrow beams are necessary for usable link budgets. On the other, beam discovery, i.e., finding the two matching beam pairs at the transmitter and the receiver, a difficult problem.

In this work, we assume a stand-alone communication model where the transmitter and the receiver are not assisted by out-of-band information regarding timing or position. The non-stand-alone model has been investigated in \cite{capone_cont2015, filipini_cont2018,capone_obst2015, marcano_macrocell2016, devoti_cont2018, abbas_context2016, xiu_outband2019, garcia_inband2018, park_locbase2017} and is not discussed in this work.
There are two main approaches to beam discovery/selection under this assumption. One, called \emph{beamsweep}, requires both the transmitter and the receiver to exhaustively search over the entire beamspace by measuring the received power for every possible transmitter-receiver beam pair.
In another, the receiver side starts listening on the channel with the widest possible beam
and step by step converges to the narrowest one. This is called
hierarchical search, see Fig. \ref{fig:bspace_hier}.


Both these techniques assume that a known signal called the \emph{synchronization signal} in cellular,
is transmitted periodically and the receiver will have to determine the direction in the beamspace where the incoming signal is stronger.
While hierarchical search in principle is superior to beamsweep in terms of search delay \cite{chiu_hier2019}, mmWave standards for both WiFi and cellular have adopted a beamsweep based procedure due to its simplicity \cite{wifi_80211ad} and \cite{3GPP_NRphyMod, 3GPP_NRphyRRC, 3GPP_NRphyProcCont, 3GPP_NRphyPhyMeas}.
In this work, we will also follow the beamsweep paradigm and our analysis and results are derived based on this assumption.

\subsection{Effect of beamforming on the beamspace}
\label{sec:beamsp}

Under the beamsweep assumption, the effective size of the beamspace
is a function of the beamforming scheme employed at the transmitter and the receiver.
That is, the beamforming scheme dictates how many directions the receiver can inspect
in a single observation. 
There are three beamforming schemes:
\begin{itemize}
\item[] \emph{Digital Beamforming:}
In this architecture, each antenna element is connected to a radio frequency (RF) chain and a pair of data converters, analog to digital or digital to analog converters (ADC or DAC). The beamforming (or spatial filtering) is performed by the digital baseband processor. For wide-band systems with a large number of antennas, this architecture can have high power consumption when high precision DACs and ADCs are used. One way to use digital beamformers at mmWave is to use DACs and ADCs with few bits of quantizer resolution.
This is attractive for beamformed systems as having digital samples form every antenna element allows the receiver to inspect theoretically infinite directions at the same time with full directional gain during only one observation of the channel. This reduction of the number of needed observations becomes significant as the beamspace grows and can potentially reduce the total power consumed during the search procedure.

\item[] \emph{Analog Beamforming:} To avoid the use of a large number of DACs and ADCs, analog beamformers perform beamforming (or spatial filtering) on the analog
(in RF or intermediate frequency) signal using RF phase shifters and power combiners (or splitters).
The use of just a pair of ADCs considerably reduces the power consumed by these front-ends and hence, these are considered a prime candidate for initial mmWave based cellular devices.
However, analog beamformers can point their beams only in one direction at a given time. This leads to potentially high delays when the beamspace is large. 

\item[] \emph{Hybrid Beamforming:} This scheme is a combination of the digital and analog beamforming.
A part of the beamforming is performed by $M$ analog RF beamformers. These beamformed signals are digitized and combined (or precoded) by the digital baseband circuitry.
This allows the receiver to inspect $M$ directions at each time instance.
Now, at an extreme $M=1$, where we have analog beamforming . At the other, $M$ equals the number of antenna elements, where the scheme is effectively digital beamforming.
Choosing $M$ trades off spatial multiplexing capabilities on the one hand, and energy consumption on the other. 
\end{itemize}

Consider a transmitter equipped with an antenna array comprised of $N_{\rm Tx}$ antenna elements
and a receiver equipped with $N_{\rm Rx}$ elements.
This means that the size of the beamspace is equal to the product of $N_{\rm Tx}$ by $N_{\rm Rx}$.
If they both use analog beamforming , then, they must visit all these directions at $N_{\rm Tx} \times N_{\rm Rx}$
separate channel inspections.
Hence, the effective size of the beamspace is, 
\begin{align}
& \text{analog:} \qquad L_{\rm an} = N_{\rm Tx} \times N_{\rm Rx}.  
\end{align}
Now, suppose that the transmitter still uses analog beamforming but the receiver uses digital.
Then, the receiver can ``listen on'' all the $N_{\rm Tx}$ directions at once and hence in this effective case the size of the beamspace becomes
\begin{align}
& \text{digital:} \qquad L_{\rm di} = N_{\rm Tx}. \quad\quad\quad
\end{align} 
Applying the same logic, for hybrid beamforming, it is easy to see that
the effective size of the beamspace is 
\begin{align}
& \text{hybrid:} \qquad L_{\rm hy} = \frac{N_{\rm Tx} \times N_{\rm Rx}}{M}
\end{align}
Other combinations of beamforming schemes on either the transmitter or the receiver yield an effective beamspace of various sizes. Adopting one beamforming scheme versus the other has a fundamental impact on the effective beamspace.
This, in turn, affects the time needed for the receiver to determine the best direction of the incoming signal. 
Furthermore, during beamsweeping the receiver FE is \emph{always on}.
An FE architecture that can listen on one or a few directions at a time will hence need to be powered on for a longer period of time to measure all the possible beam pairs. This can mean a considerable increase in the effective power consumed by analog and hybrid beamformers.
In the next sections we quantify the impact of the size of the beamspace
and the choice of the beamformers on the energy consumed by beam discovery procedure.

\subsection{The 3GPP NR paradigm}
\label{sec:sigsysmodgpp_par}

\begin{figure}
\begin{center}

 \includegraphics[width=0.95\linewidth]{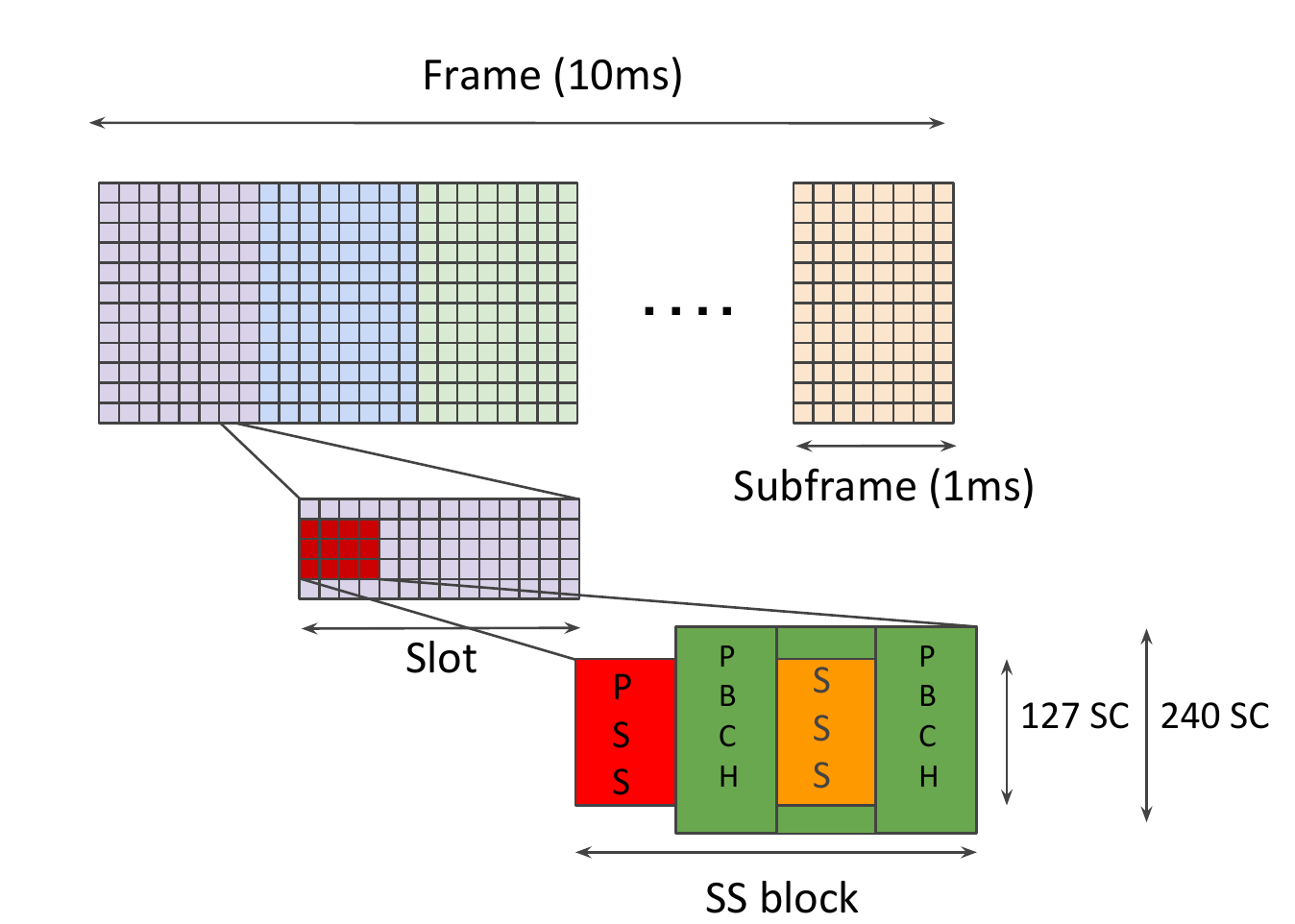}
\end{center}
\caption{Frame structure of 3GPP NR. A frame is divided into 10 subframes, a subframe into 8 slots and a slot into 14 OFDM symbols. A single SS block takes up 4 OFDM symbols.
}
\label{fig:nr_frame}    
\end{figure}

As an illustrative example of a system using beamsweep, we will discuss the 3GPP NR physical layer standard for initial access (IA).
Our analysis and results in the proceeding sections will all have an NR system as an underlying assumption.
We chose NR for two reasons.
One, it is the standard defined for 5G cellular systems so it is expected to be adopted by millions of devices
in the coming years. Two, the NR standard offers a well-defined set of assumptions regarding the 
beam discovery process and time and frequency numerology.
This allows us to evaluate the different beamforming schemes with respect to energy consumption within a widely accepted context.

We next present the beam discovery process which is one of the basic steps of NR \emph{initial access}, i.e.,
the procedure of establishing a link-layer connection between a base-station, referred to as
NR 5G nodeB (gNB) in the NR standard, and a mobile device or user equipment (UE).

\subsubsection{NR beam discovery}
\label{sec:nr_beamdis}


Initial access involves a set of message exchanges between the gNB and the user equipment, whereby the user equipment identifies a serving gNB and synchronizes to it.
The user equipment learns the physical cell identity of the gNB,
sends back its own ID and finally attaches to the cell.
In this work we will focus only on the first part of NR initial access: beam and cell discovery on the user equipment side, since this is the most critical in terms of detection delay and energy consumption. 

On the gNB side, all the $N_{\rm Tx}$ directions are swept with a periodicity of $T = 20~\text{ms}$. 
More specifically, every $T$ and for an interval of duration $T_{\rm ssb} = 5~\text{ms}$,
the gNB transmits $B$ blocks of four OFDM symbols in $N_{\rm TX}$ directions.
These transmissions during $T_{\rm ssb}$ are called a synchronization signals (SS) burst
and each block of the four OFDM symbols, an SS block.
Each SS block is comprised of the primary synchronization signal (PSS), the secondary synchronization signal (SSS) and
the physical broadcast channel (PBCH).
The PSS and the SSS together make up the physical ID of the cell.
The first one takes one of three possible values among $\{ 0,1,2\}$ while the latter, one of $\{ 0,1, \ldots, 335\}$,
resulting in a total of $1008$ unique cell IDs. 
Each of these signals takes up 127 sub-carriers in frequency,
while an entire SS block, including PBCH, $240$ sub-carriers.
In Fig. \ref{fig:nr_frame} we depict the time-frequency resources occupied by the SS block within an NR frame, subframe and slot. 

\begin{figure}
\begin{center}

 \includegraphics[width=0.95\linewidth]{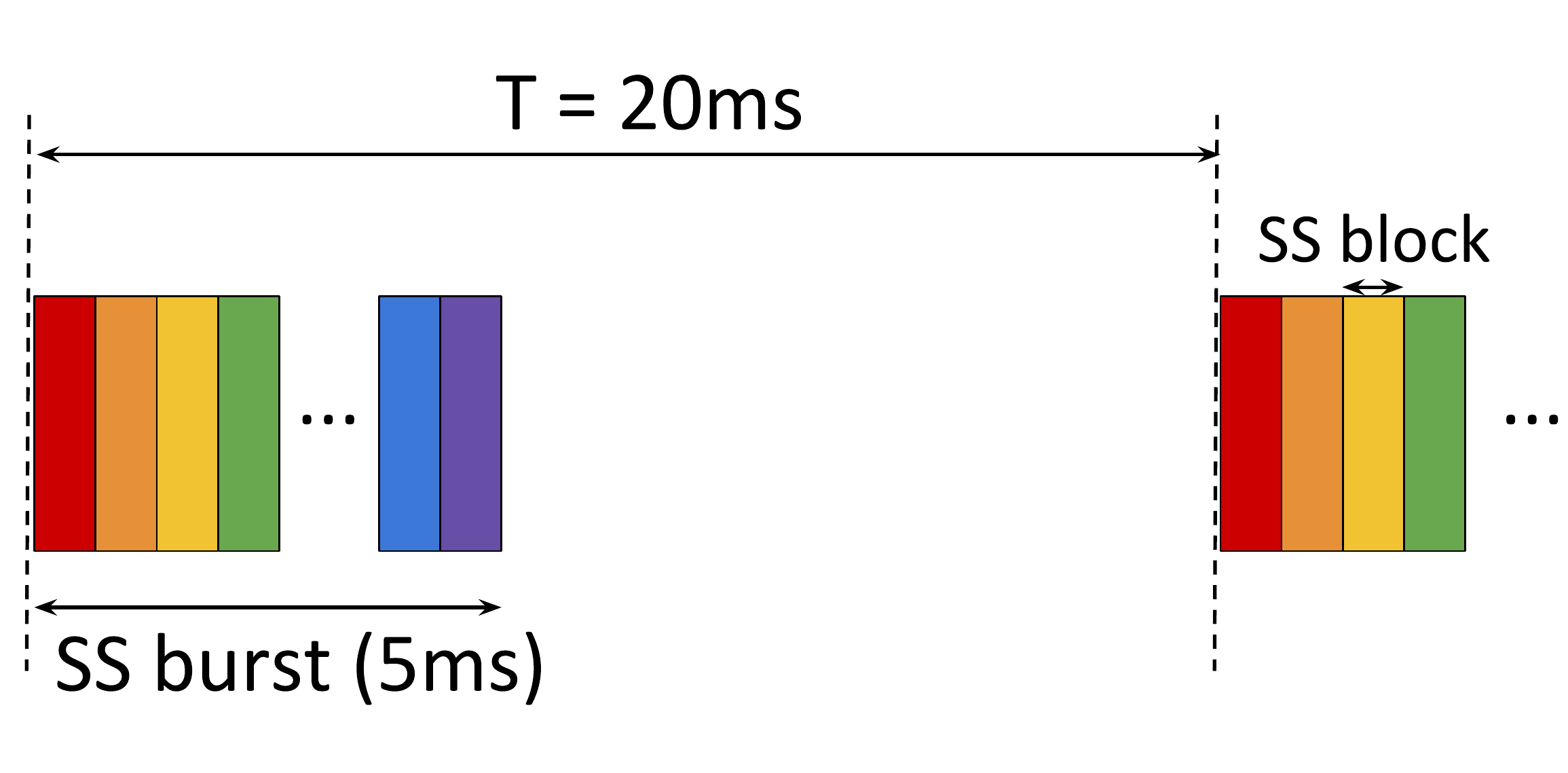}
\end{center}
\caption{SS burst. Each burst lasts for $5 {\rm ms}$ and is repeated every $20 {\rm ms}$. Each SS block within an SS burst is shown in a different color.}
\label{fig:ss_burst}    
\end{figure}

\begin{figure}
\begin{center}

 \includegraphics[width=0.95\linewidth]{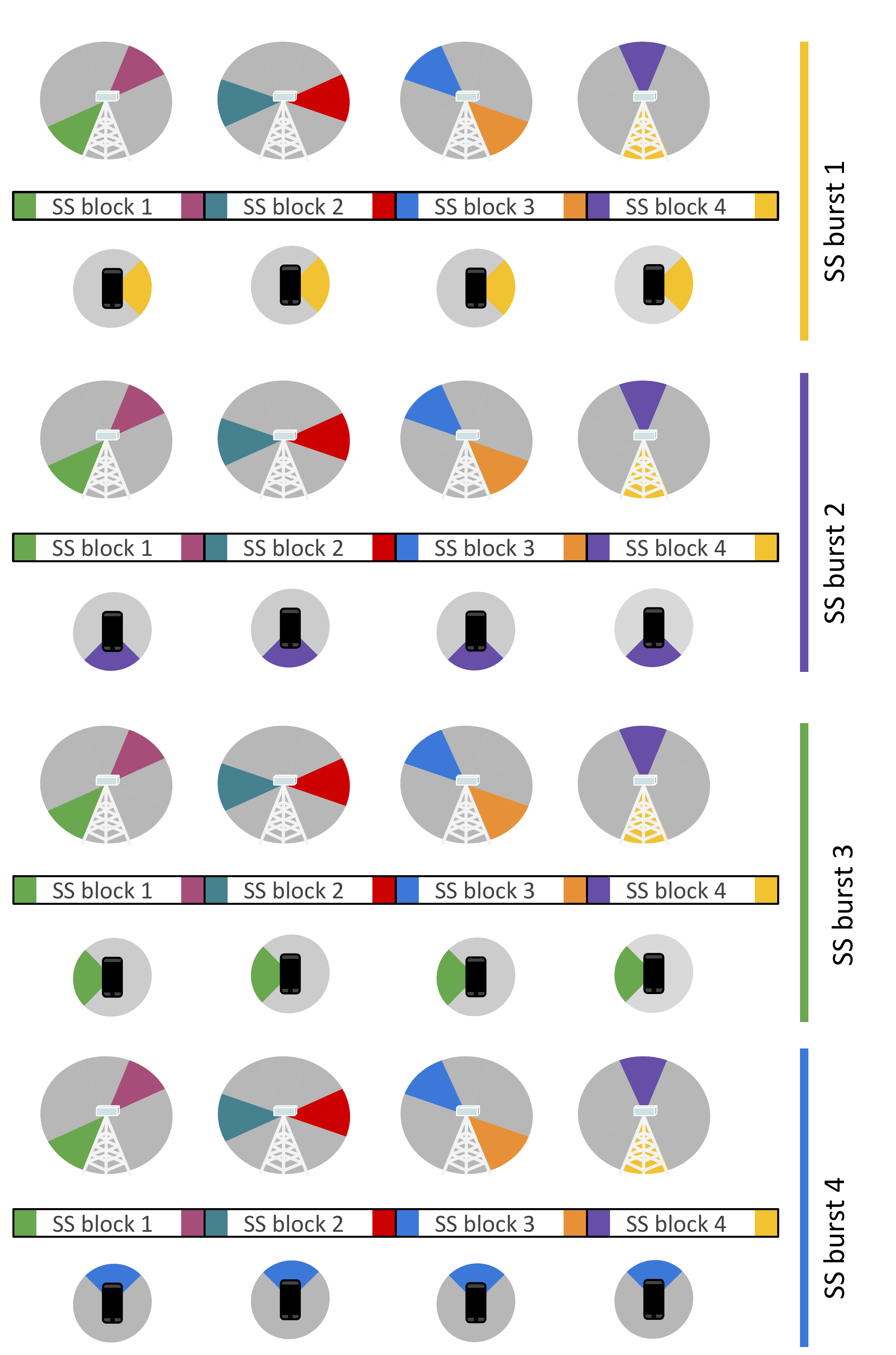}
\end{center}
\caption{Example of beam discovery within four SS bursts. During each SS burst, the gNB covers $N_{\rm Tx} = 8$ directions in $4$ SS blocks. The user equipment on the other hand scans all the $N_{\rm Rx} = 4$ directions in four SS bursts. Thus, the user equipment and the gNB together need four SS bursts to cover the beamspace of size $L = 16$.}
\label{fig:ss_burst_xmpl}    
\end{figure}

Fig. \ref{fig:ss_burst} shows an example of several SS blocks within a single SS burst.
For the mmWave bands, the NR standard provisions a total of $B=32$ SS blocks during each SS burst.
It is assumed that each SS block is transmitted simultaneously in two directions using hybrid beamforming.
Hence, a mmWave gNB can support up to 64 non-overlapping directions.
As an example, in Fig. \ref{fig:ss_burst_xmpl}, we depict a scenario of a gNB beamsweeping $N_{\rm Tx} = 8$
directions using $B = 4$ SS blocks per SS bursts. 

Now, the user equipment is also sweeping $N_{\rm Rx}$ directions in searching for SS burst , see Fig. \ref{fig:ss_burst_xmpl}.
These signals are known to the user equipment. It has to determine which one of the possible three
PSS sequences and 336 SSS sequences were sent.
Since the structure of an SS block is also known, once the PSS is detected and the optimal direction is found, the user equipment
will move on to detecting the SSS within the same SS block which is the next to the next OFDM symbol,
as shown in Fig. \ref{fig:nr_frame}.
Thus, the most critical part of beamsweeping is the detection of the PSS which unlocks all the remaining steps of initial access.

The user equipment is assumed to use analog beamforming.
Hence, unlike the gNB it can probe only one direction at a time. 
Since the gNB sends the SS blocks in two directions simultaneously,
according to Sec. \ref{sec:beamsp} the effective size of the beamspace around the gNB and the user equipment
is $L = \left({N_{\rm TX} \times N_{\rm RX}}\right)/{2}$. 

\section{Signal and System Model}
\label{sec:sigsysmod}

We will leverage the NR beam discovery process described earlier for our analysis and modeling,
with a few simplifying assumptions.
We consider a single cell of radius $d_{r}$ with the gNB situated at the center transmitting the PSS signals periodically. 
The user equipment, through analog beamforming  will sweep $N_{\rm Rx}$ directions sequentially in search of the PSS signal.

Our first simplification is assuming analog beamforming at the gNB. 
Therefore, the user equipment and the gNB together will have to sample a beamspace of size $L_{\rm ang} = N_{\rm tx} \times N_{\rm rx}$.
Note that for user equipment at low SNR, i.e., those at the edge of the cell, it may be needed to cycle through the beamspace more than once. We will denote each cycle with $k = 1, 2, \ldots,K$.

Our next assumption is that, similar to \cite{barati2016initial},
the dominant path between the gNB and the user equipment is a line of sight (LOS) path aligned with exactly one of the transmitter-RX directions/sectors $\ell^{\rm *}$ in the beamspace.
Although this assumption may seem unrealistic, since real channels
are seldom comprised of a single path, it is only used to derive our detectors below. 
In our evaluation and simulations, we will test our detectors using a channel model derived from real measurements \cite{AkdenizCapacity:14}.

With these assumptions in mind, we model the received post-analog-BF received complex signal $\ybf_{k\ell}$
at the user equipment during the $\ell$-th sampling of the beamspace
and $k$-th sweeping cycle as:
\beq{}
\label{eq:rx_sig}
\ybf_{k \ell} = \ubf_{k \ell}^{*}\Hbf_{k \ell} \vbf_{k \ell}^{*} \xbf + \nbf_{k \ell}, \quad
  \nbf_{k\ell } \sim {\mathcal CN}(0,\sigma_{k\ell}^2\Ibf_D). 
\eeq{}

The transmitted PSS signal $\xbf \in \C^D$ is in a signal space with
$D \approx T_{\rm PSS} \times W_{\rm pss}$ orthogonal degrees of freedom,
where $T_{\rm PSS}$ and $W_{\rm pss}$ are the PSS duration in time and bandwidth occupied by the PSS signal, respectively.
We assume the PSS signal to be unit norm, $\Vert \xbf \Vert = 1$.
The vectors $\ubf_{k \ell} \in \C^{N_{\rm RX}}$ and $\vbf_{k \ell} \in \C^{N_{\rm TX}}$ are, respectively, the user equipment and gNB side beamforming vectors 
along a transmitter-receiver direction $\ell$ in the sectorized beamspace at the $k$-th sweep cycle.
They are assumed to be of fixed-norm. 

The MIMO channel $\Hbf$ is assumed to be flat-fading within the PSS bandwidth and constant over the PSS transmission time $T_{\rm PSS}$. 
It is defined as
\beq{}
\label{eq:chan}
\Hbf = h_{k \ell}\ubf_{\ell^\text{*}} \vbf_{\ell^\text{*}}^*, 
\eeq{}
where $h_{k \ell}$ is a small-scale fading coefficient.
The vectors $\ubf_{\ell^\text{*}}$ and $\vbf_{\ell^\text{*}}$ are the
spatial signatures of the user equipment and the gNB antenna arrays 
describing a single LOS path between them,
aligned with only one of the transmitter-RX directions in the beamspace.
They are given as 

\beq{}
\label{eq:sp_sig}
  \begin{array}{ll}
  \ubf_{\ell^\text{*}} = [1, e^{-j 2 \pi \Delta \cos (\phi_{\ell^\text{*}}^{\rm Rx} )}, \hdots, e^{-j 2 \pi \Delta (N_{\rm RX} -1)\cos (\phi_{\ell^\text{*}}^{\rm Rx} )} ]^T \\
  \vbf_{\ell^\text{*}} = [1, e^{-j 2 \pi \Delta \cos (\phi_{\ell^\text{*}}^{\rm Tx} )}, \hdots, e^{-j 2 \pi \Delta (N_{\rm TX} -1)\cos (\phi_{\ell^\text{*}}^{\rm Tx} )} ]^T,
  \end{array}
\eeq{}
where $\phi_{\ell^*}^{\rm Rx}$ and $\phi_{\ell^*}^{\rm Tx}$ are the angles of arrival and departure in the directions of $\ell^\text{*}$,
and $\Delta$ is the distance between elements of the antenna arrays measured in wavelengths.
Finally, $\nbf$ in \eqref{eq:rx_sig} is the i.i.d. complex additive white Gaussian noise vector
with co-variance $\sigma_{k\ell}^2\Ibf_D$, and $\Ibf_D$ is the $D\times D$ identity matrix. 

The aim of the user equipment and the gNB is to mutually steer their beams along their spatial signatures defined in \eqref{eq:sp_sig}.
That is, to apply beamforming vectors, $\ubf_{\ell}$ and
$\vbf_{\ell}$, as close to 
to $\ubf_{\ell^\text{*}}$ and $\vbf_{\ell^\text{*}}$ as possible.
This, based on our models in eq. \eqref{eq:rx_sig} and
eq. \eqref{eq:chan} and our fixed-norm assumption over the beamforming vectors, 
leads to the maximization of the energy of the received signal vector $\ybf_{k \ell}$. 


\subsection{Signal Detection}
\label{sec:sigdetect}
Given the signal and transmission model above we now derive
the PSS detector through hypothesis testing.
We will follow the same line of analysis given in \cite{barati2016initial}.
Since we have assumed that the dominant path from the gNB to the user equipment is along
a single transmitter-receiver beamspace direction $\ell^\text{*}$ (here the asterisk denotes the true or best direction), 
for each probed direction in each sweeping cycle
our null hypothesis, $H_0$, is that the signal is not present.
Conversely, our alternative hypothesis, $H_1$, is that the signal is present, i.e., the user equipment is probing the direction in which the gNB is transmitting. 
Hence, eq. \eqref{eq:rx_sig} under $H_0$ (signal not present) and $H_1$ (signal present at correct beamspace direction) becomes

\beq
\label{eq:hyp}
\begin{array}{ll}
  H_0: ~ \ybf_{k \ell} = \nbf_{k \ell}\\
  H_1: ~ \ybf_{k \ell} = \alpha_{k \ell^\text{*}} \xbf + \nbf_{k \ell^\text{*}},
  \end{array}
\eeq

where the scalar channel coefficient $\alpha_{k \ell^\text{*}}$ is the result of applying the beamforming vectors
along the spacial signatures of the user equipment and the gNB and is given by
\beq{}
\alpha_{k \ell^\text{*}} = \ubf_{\ell^\text{*}}^* \Hbf_{k \ell^\text{*}} \vbf_{\ell^\text{*}} = 
h_{k \ell^\text{*}} \ubf_{\ell^\text{*}}^* \ubf_{\ell^\text{*}} \vbf_{\ell^\text{*}}^* \vbf_{\ell^\text{*}}.
\eeq{}
We define the probability density of the received signal $\ybf$ under the two hypotheses as
$p(\ybf | H_i, \alphabf, \bf{\sigma^2}, \ell^\text{*} )$ for $i = \{0,1\}$,
where the set $\ybf$ contains all the observed signals $\ybf_{k \ell}$ in $K$ beam sweeps.
The model contains several unknown parameters, namely, $\alphabf$ the set of all the
channel coefficients $\alpha_{k \ell^\text{*}}$, and ${\bf \sigma}^2$ the set of all noise power levels. 

Due to these unknown parameters, we will rely on the widely used
generalized likelihood ratio test (GLRT) method \cite{VanTrees:01a}.
The GLRT takes the likelihood distribution under each hypothesis maximized with respect to the unknown parameters and compares their ratio to a threshold.
Alternatively, we can use the minimum log-likelihoods of the signal distribution under each hypothesis. 
These are given as
\begin{subequations}
\label{eq:MLhyp}
\beqa
  && \Lambda_0 := \min_{\sigmabf^2} -\ln p(\ybf|H_0, \sigmabf^2) \\
  && \Lambda_1 := \min_{\sigmabf^2,\alphabf} -
  \ln p(\ybf|H_1, \sigmabf^2,\alphabf).
\eeqa
\end{subequations}
We then use the test 
\beq{}
\label{eq:log_test}
\Lambda := \Lambda_0 - \Lambda_1 \underset{H_0}{\overset{H_1}{\gtrless}} \; \theta
\eeq{}
for a threshold $\theta$. We observe that under the assumption that the noise vectors are independent
in different measurements,
\begin{subequations}
\label{eq:glrt_sum}
\beqa
  && \ln p(\ybf|H_0, \sigmabf^2,\ell^\text{*}) = \sum_{k=1}^K \sum_{\ell=1}^L \ln p(\ybf_{k\ell}|\sigma_{k \ell}^2,\ell^\text{*})\\
  && \ln p(\ybf|H_1, \sigmabf^2,\alphabf,\ell^\text{*}) = \nonumber \\
  && \sum_{k=1}^K \sum_{\ell=1}^L \ln p(\ybf_{k\ell}|\sigma_{k \ell}^2,\alpha_{k \ell^\text{*}},\ell^\text{*}).
\eeqa
\end{subequations}
Therefore, the negative log-likelihoods in eq.\eqref{eq:MLhyp} can be re-written as
\begin{subequations}
\label{eq:MLhyp2}
\beqa
  && \Lambda_0 =  \sum_{k=1}^K \sum_{\ell=1}^L \Lambda^{k\ell}_0(\ell^\text{*}). \\
  && \Lambda_1 = \sum_{k=1}^K \sum_{\ell=1}^L \Lambda^{k\ell}_1(\ell^\text{*}),
\eeqa
\end{subequations}

where $\Lambda^{k\ell}_0(\ell^\text{*})$ and $\Lambda^{k\ell}_1(\ell^\text{*})$
are the minimum negative log-likelihoods in each measurement and given as:
\begin{subequations}
\label{eq:MLhyp_local}
\beqa
  && \Lambda^{k\ell}_0(\ell^\text{*}) = \min_{\sigma^2} -\ln p(\ybf_{k\ell}|\sigma_{k \ell}^2,\ell^\text{*}). \\
  && \Lambda^{k\ell}_1(\ell^\text{*}) = \min_{\sigma^2,\alpha_{k \ell}} -\ln p(\ybf_{k\ell}|\sigma_{k \ell}^2,\alpha_{k \ell}, \ell^\text{*}).
\eeqa
\end{subequations}

Since the received signal in our model, given by eq. \eqref{eq:hyp}, is Gaussian conditional on the parameters, hence
\begin{align}
\label{eq:ln_local_h0}
  { -\ln p(\ybf_{k \ell}|\sigma_{k \ell}^2,\ell^\text{*}) } 
  & =
    \frac{1}{\sigma_{k \ell}^2} \|\ybf_{k \ell}\|^2
    + D\ln(\pi\sigma_{k\ell}^2)
\end{align}
under the $H_0$ hypothesis, and 
\begin{align}
\label{eq:ln_local_h1}
  { -\ln p(\ybf_{k \ell}|\sigma_{k \ell}^2,\alpha_{k \ell^\text{*}},\ell^\text{*}) }
  & =
    \frac{1}{\sigma_{k \ell}^2} \|\ybf_{k \ell} - \alpha_{k \ell^\text{*}} \xbf\|^2
    + D\ln(\pi\sigma_{k\ell}^2)
\end{align}
under the $H_1$ hypothesis.
Now, we use the above eqns. (\ref{eq:ln_local_h0}) and (\ref{eq:ln_local_h1}), to obtain estimates of the unknown variables which are their minimizers. 

We first take $\Lambda^{k\ell}_1(\ell^\text{*})$.
The channel coefficient $\alpha_{k \ell^\text{*}}$ that minimizes the log negative likelihood when the signal is present is given as 
\beq{}
\label{eq:alpha_1}
  \alphahat_{k \ell^\text{*}} = \frac{\xbf^*\ybf_{k\ell}}{\|\xbf\|^2}.
\eeq{}
Next we move on to minimize over the variance $\sigma^2$, which occurs at
\beq
\label{eq:sigma_1}
  \sigmahat_{k \ell}^2 = \frac{1}{D} \left( \|\ybf_{k\ell}\|^2 - \frac{|\xbf^*\ybf_{k\ell}|^2}{\|\xbf\|^2} \right).
\eeq
When the signal is not present, i.e., $\ell \neq \ell^\text{*}$,
we minimize $\Lambda^{k\ell}_0(\ell^\text{*})$ over the variance $\sigma^2$. 
The minimum is obtained at 
\beq{}
\label{eq:sigma_0}
\sigmahat_{k\ell}^2 = \frac{\|\ybf_{k\ell}\|^2}{D}.
\eeq{}

Substituting eq.\eqref{eq:alpha_1}, eq.\eqref{eq:sigma_1} and eq. \eqref{eq:sigma_0}, in eq. \eqref{eq:MLhyp_local} we obtain

\begin{subequations}
\label{eq:MLhyp_local_last}
\beqa
  && \Lambda^{k\ell}_0(\ell^\text{*}) = D \ln ( \frac{ \pi e}{D} \|\ybf_{k\ell}\|^2), \\
  && \Lambda^{k\ell}_1(\ell^\text{*}) = D \ln \left( \frac{ \pi e}{D} \left(\|\ybf_{k \ell} \|^2 - \frac{| \xbf^* \ybf_{k \ell}|^2}{\| \xbf\|^2}\right) \right).
\eeqa
\end{subequations}
Combining \eqref{eq:MLhyp_local_last} and from the test \eqref{eq:log_test}, and eq. \eqref{eq:MLhyp2}
we have the following log-likelihood difference
\beq
\label{eq:log_dif}
  \Lambda =
    \sum_k^K \sum_{\ell}^L - D \ln \left( 1 - \rho_{k \ell} \right),
\eeq
where the $\rho_{k \ell}$ is the normalized energy of the correlation of the
known signal with the received signal, obtained as
\[
  \rho_{k \ell} = \frac{|\xbf^* \ybf_{k \ell}|^2}{\|\xbf\|^2 \|\ybf_{k \ell}\|^2}.
\]
This, not surprisingly, means that while the user equipment probes each direction in $K$ beamsweeps,
it should correlate the incoming signal with the local replica of the known signal.
Then, since eq. \eqref{eq:log_dif} is an increasing function of $\rho_{k \ell}$,
and due to our assumption that the signal is present only in one direction within the beamspace,
the best direction $\ellhat^\text{*}$ is the one in which the correlation was the highest, that is,
\beq
\ellhat^\text{*} = \argmin_{\ell = 1, ..., L} \sum_k^K\sum_\ell^L \ln(1 - \rho_{k \ell}). 
\eeq

While the above detector was derived for analog beamforming under favorable beamspace assumptions,
it can also be used for both hybrid and digital beamforming .
The difference is that with these beamforming architectures, the receiver has access to
$N_{RF} >1$ RF chains and can essentially operate as $N_{RF}$ parallel analog systems.
In the digital beamforming case $N_{RF}$ is equal to $N_{\rm Rx}$. 
Note that with a fully digital detector the receiver can test all the angles,
including those not perfectly aligned in the beamspace.
However, for the sake of simplicity and better comparison with analog beamforming ,
we will not consider a more powerful detector but assume that in digital beamforming too
the arrival angles are aligned with the beamspace directions.

\subsection{RF Architectures and power consumption}
\label{sec:rf_arch}
While standards documents on the implementation of mmWave cellular systems assume user equipments
with analog beamforming and only a single RF chain, we believe that there is
much value in using fully digital beamforming .
To make our case more concrete in terms of energy consumption we give here an overview of
the receiver's mmWave front-end. 
We model the power consumption of analog, hybrid and fully digital
and fully digital with low-resolution quantization front tends.

\subsubsection{mmWave front-ends }
\label{sec:mmw_fe}
\begin{figure}
  \centering
  \includegraphics[page=1,trim={1cm 4cm 1cm 1cm}, clip, width=0.5\textwidth]{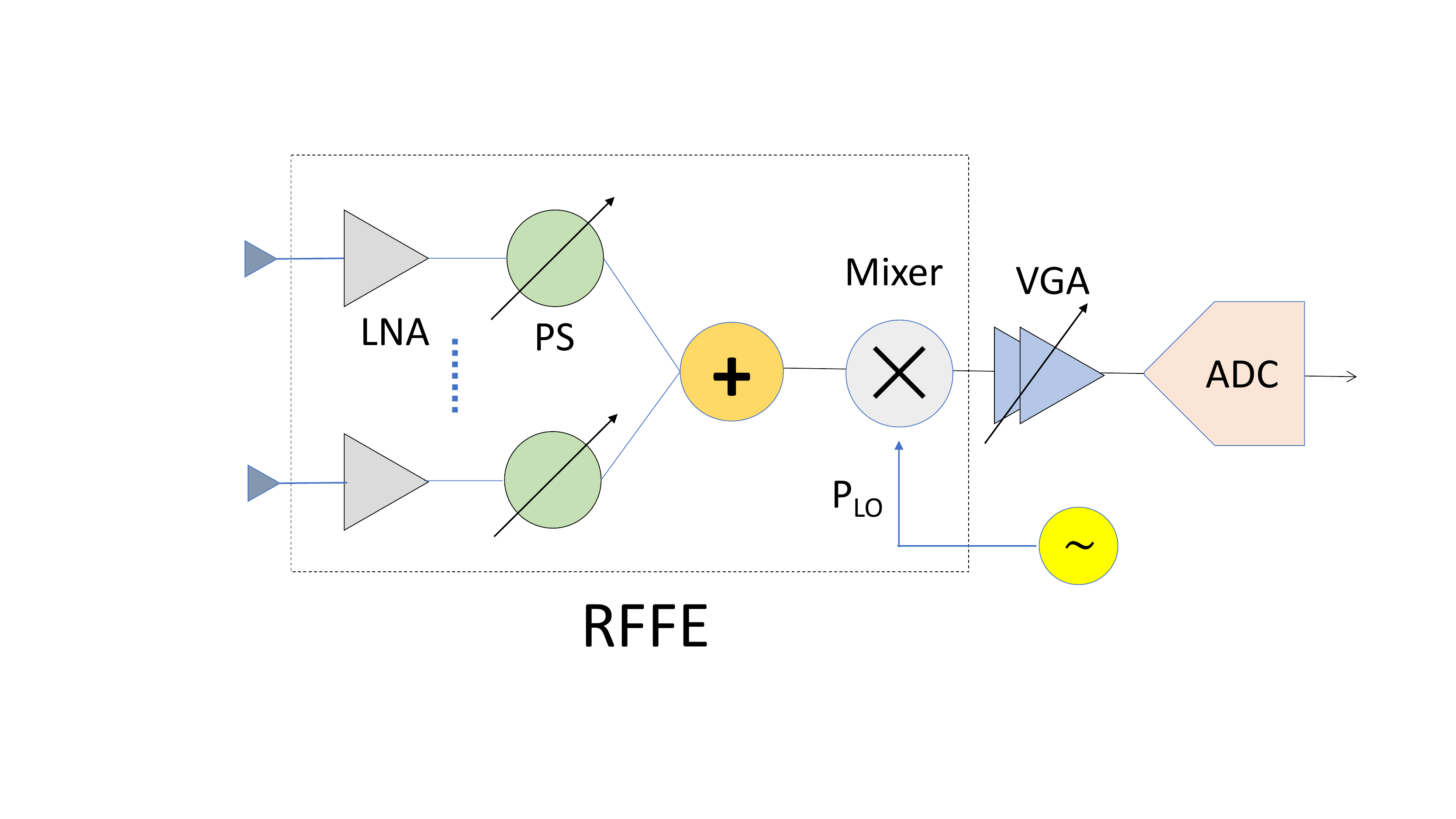}
  \caption{Analog beamforming receiver architecture.}
  \label{fig:analogRxArch}
\end{figure}

\paragraph{Analog front-end:} Consider the mmWave receiver's analog front-end shown in Fig. \ref{fig:analogRxArch}.
It is comprised of $N_{\rm RX}$ low noise amplifiers (LNA)s,
$N_{\rm RX}$ phase shifter (PS), a mixer, a combiner and a pair of analog to digital converter (ADC). 
The D.C. power consumption of each LNA, $P_{\rm DC}^{\rm LNA}$ is a function of its gain $G_{\rm LNA}^{\rm PS}$, the noise figure $N_{\rm LNA}$
and the figure of merit (FoM) and is given as \cite{dutta2019twc, Song2008, Adabi07} 

\beq
\label{eq:lna_analog}
P_{\rm LNA}^{\rm DC} = \frac{G_{\rm LNA}^{PS}}{\text{FoM} (N_{LNA} - 1)}, 
\eeq
in linear scale.

\begin{figure}
  \centering
  \includegraphics[page=2,trim={1cm 3.5cm 1cm 1cm}, clip, width=0.5\textwidth]{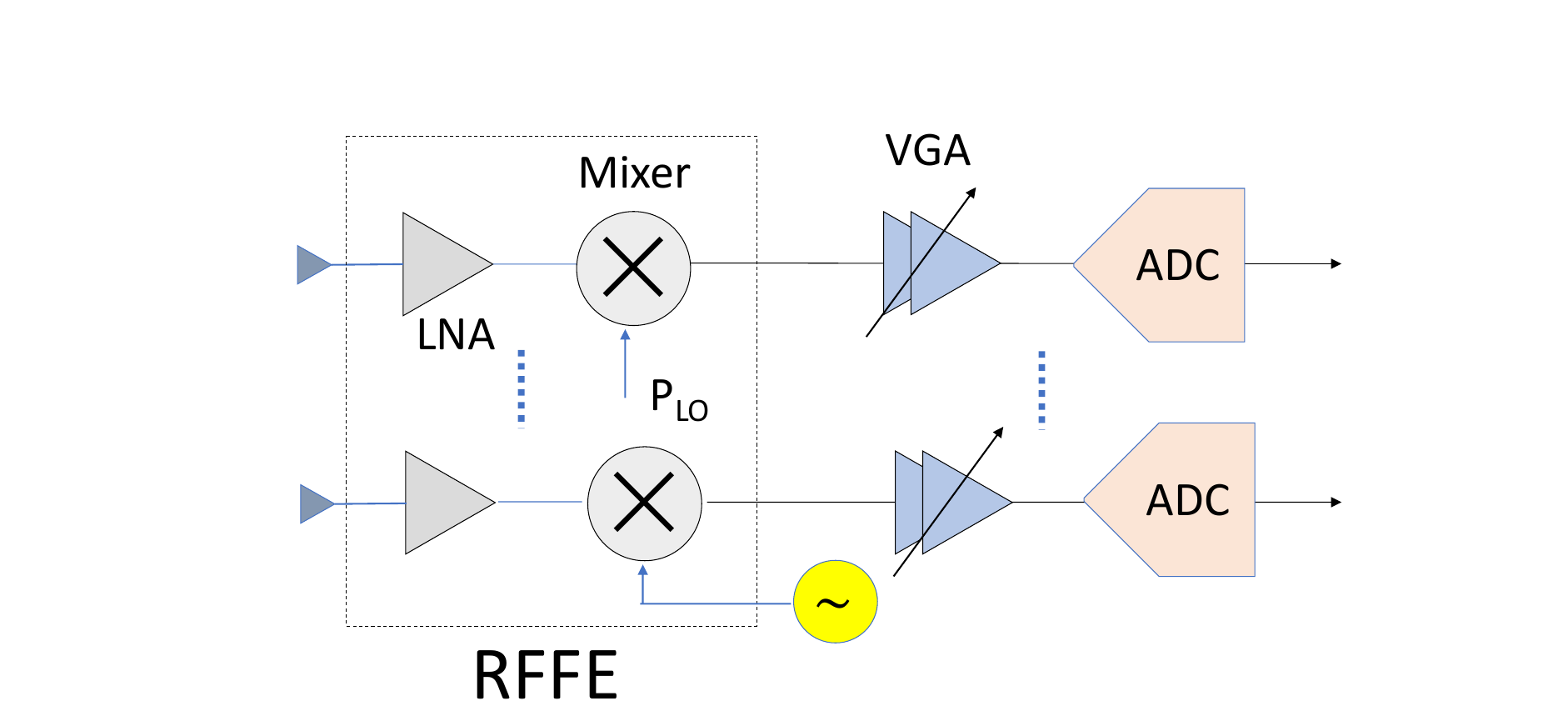}
  \caption{Digital beamforming receiver architecture.}
  \label{fig:digitalRxArch}
\end{figure}

The PS and mixer are considered to be passive elements.
While they do not burn any power, they do however introduce insertion losses (IL),
which need to be compensated for. 
For example, the LNA gain above needs to be high enough to offset the loss introduced by the phase shifter $IL_{\rm PS}$.
That is, if an LNA in a circuit without any PS had a gain $G_{\rm LNA}$,
then $G_{\rm LNA}^{\rm PS} = G_{\rm LNA} + IL_{\rm PS}$.

For the ADCs, the power consumption is a function of the sampling frequency $F_{\rm s}$, ADC's 
figure of merit and the resolution in bits:
\beq{}
\label{eq:adc_pow}
P_{\rm ADC} = \text{FoM} \times F_{\rm s} \times 2^q,
\eeq{}
where $q$ is the number of resolution bits.
To keep the variable incoming signal within the ADCs dynamic range
while keeping a constant baseband power $P_{\rm BB}^{\rm out}$ it is necessary
to apply gain control to the input of the ADC. 
This is performed by the variable gain amplifier (VGA) with a gain range between $0$ and $G_{\rm VGA}^{\rm max}$.
For the analog beamforming case this maximum gain is calculated as:
\beqa
\label{eq:vga_gain}
&&G_{VGA}^{max} = \nonumber \\
&&P_{\rm BB}^{\rm out} - 10\log(N_{\rm RX}) + IL_{\rm mix} \nonumber \\
&& - (G_{\rm LNA}^{\rm PS} - IL_{\rm PS}) - P_{\rm RX}(d = d_{\rm max}),
\eeqa
where $IL_{\rm mix}$ is the insertion loss introduced by the mixer and $P_{\rm RX}(d = d_{\rm max}$ is the
maximum distance between the transmitter and the receive, i.e., the cell edge.

Given the above gain, the D.C. power draw of the VGA is given as 
\beq
\label{eq:p_vga}
P_{\rm VGA}^{\rm DC} = \frac{G_{\rm VGA}^{\rm max} F_{\rm s}}{\text{FoM} A_{\rm chip}},
\eeq
Where FoM is the FoM of the VGA, $F_{\rm s}$ the sampling bandwidth in ${\rm GHz}$ and $A_{\rm chip}$ is the active area in mm$^2$.

Based on the above, the total power consumption of an analog mmWave front-end can be calculated as

\beq{}
\label{eq:an_pow}
P_{\rm an} = N_{\rm RX} P_{\rm LNA}^{\rm DC} + P_{\rm LO} + P_{\rm ADC} + P_{\rm VGA}^{\rm DC},
\eeq
where $P_{\rm LO}$ is the power draw of the local oscillator.

\paragraph{Digital front-end:} Consider the digital front-end in Fig. \ref{fig:digitalRxArch}.
Since we have introduced most of the components in the analog front-end case,
computing the power for the digital case much easier. 
Notice in the figure, that since beamforming is done in baseband there is no need for phase shifters.
Neither do the LNA gains need to compensate for their insertion losses.
Therefore, we will use $G_{\rm LNA}$ instead of $G_{\rm LNA}^{\rm PS}$. 
However, the number of the ADCs is increased to $N_{\rm RX}$, so is the
number of VGAs, mixers and local oscillators. 
Thus, the DC power draw by the LNA and VGA are given as
\beq
\label{eq:lna_dig}
P_{\rm LNA}^{\rm DC} = \frac{G_{\rm LNA}}{\text{FoM} (N_{\rm LNA} - 1)}, 
\eeq
and 
\beqa
\label{eq:vga_gig}
G_{\rm VGA_{dig}}^{\rm max} = \nonumber \\
&& P_{\rm BB}^{\rm out} - 10\log(N_{\rm RX}) + IL_{\rm mix} \nonumber \\
&& - (G_{\rm LNA}) - P_{\rm RX}(d = d_{\rm max}).
\eeqa
With these, the total power consumed by a fully digital mmWave front-end is
\beq{}
\label{eq:dig_pow}
P_{\rm dig} = N_{\rm RX} (P_{\rm LNA}^{\rm DC} +P_{\rm LO} + P_{\rm ADC} + P_{VGA_{\rm dig}}^{\rm DC}).
\eeq

\paragraph{Digital front-end with low quantization ADCs:}
The front-end here remains the same as in Fig. \ref{fig:digitalRxArch}. 
The only thing that changes is the resolution $q$ of the ADCs. 
Notice in eq. \eqref{eq:adc_pow}, that the power consumption is exponential in $q$.
Since the power consumption of a fully digital mmWave front-end grows linearly
in the number of antenna elements $N_{\rm RX}$ as given in eq. \eqref{eq:dig_pow}, reducing the resolution
is the only meaningful way of bringing the power draw of a fully digital front-end. 
In our calculations below and in Section \ref{sec:evsim} we will show how
reducing the quantization resolution affects both the power draw of the front-end
and the energy consumption during the beam discovery process. 

\begin{figure}
  \centering
  \includegraphics[page=4,trim={1cm 2.3cm 1cm 1cm}, clip, width=0.5\textwidth]{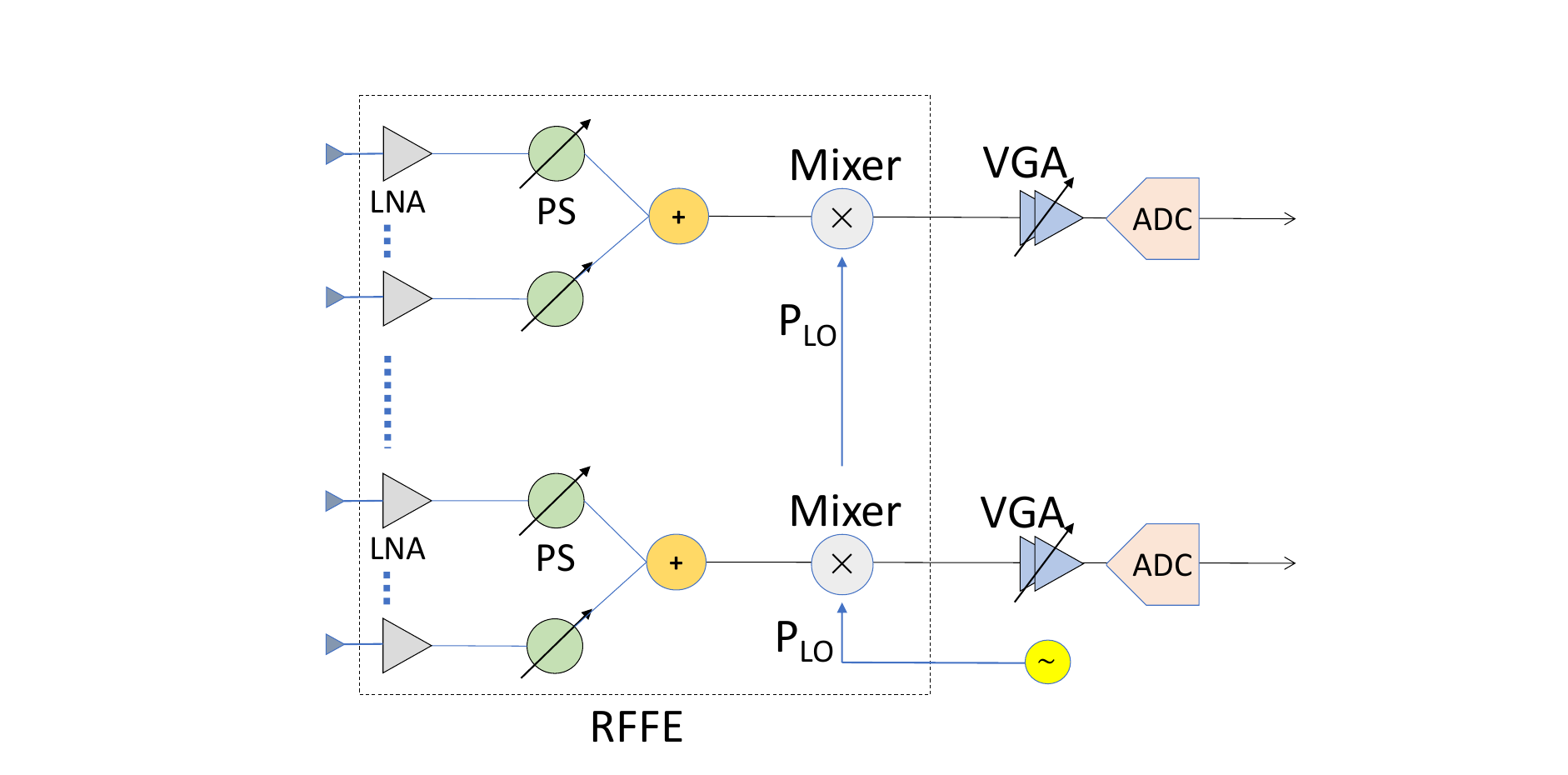}
  \caption{Sub-array hybrid beamforming receiver architecture.}
  \label{fig:hybridRxArch}
\end{figure}

\paragraph{Hybrid front-end:} For the sake of completeness, we also present
the mmWave hybrid front-end in Fig.\ref{fig:hybridRxArch}. 
There are a few options in designing
a hybrid front-end.
Here, we show the ``industry-standard'' sub-array architecture where
the antenna array is divided by the
number of supported digital streams. 
The depicted front-end can support $M = 2$ digital streams each connected to sub-arrays of size $N_{rm RX}/2$.
Notice that this circuit has elements from both the fully digital and analog front-ends:
$N_{\rm RX}$ phase shifters and $M = 2$ pairs of ADCs.
Thus, the total power consumption of the hybrid front-end is

\beq{}
\label{eq:hy_pow}
P_{\rm hy} = N_{\rm RX} P_{\rm LNA}^{\rm DC} + M (P_{\rm LO} + P_{\rm ADC} + P_{\rm VGA}^{\rm DC}).
\eeq

\subsubsection{front-end power consumption}
\label{sec:fr_powcons}

\begin{table*}
\begin{center}
\caption{mmWave Receiver's Power Consumption (mW). A receiver array size of 16 antennas
and a sampling rate of $1 \text{GHz}$ are assumed.}
\label{tab:1}    
\begin{tabular}{llllll}
\hline\noalign{\smallskip}
{\bf front-end Architecture} & {\bf RFFE} & {\bf VGA} & {\bf ADC ($q = 10$)} & {\bf ADC ($q = 4$)}& {\bf Total} \\
\noalign{\smallskip}\hline\noalign{\smallskip}
Analog & 257.3 & 1.55 & 133.12 & -- & 391.97\\
Hybrid ($M = 2$) & 267.3 & 3.1 & 266.24 & -- & 536.64\\
Digital (High res.) & 184.7 & 24.8 & 2129.9 & -- & 2339.4\\
Digital (Low res.) & 184.7 & 24.8 & -- & 33.28 & 242.78\\

\noalign{\smallskip}\hline
\end{tabular}
\end{center}
\end{table*}

To give a better picture of each front-end's power consumption,
we will now give numerical examples.
We will use power draws and losses of individual components as reported in 
studies in the literature and give a total number for each front-end when put together.

For an LNA with gain $G_{\rm LNA} = 10~\text{dB}$ with a noise figure of
$N_{\rm LNA} = 3~\text{dB}$ and $\text{FoM} = 6.5 \text{mW}^{-1}$ 
we have used the $90~\text{nm}$ CMOS LNA reported in \cite{dutta2019twc, Song2008, Adabi07}.
We have assumed a PS insertion loss $IL_{PS} = 10~\text{dB}$ which the
analog front-end LNA gain $G_{\rm LNA}^{\rm PS}$ needs to compensate for. 

For the ADC, we have used a 4-bit Flash-based ADC reported in \cite{Nasri2017}
with an $\text{FoM}$ of $67.6 \text{fJ/conversion step}$. 
While the 4-bit ADC falls into the category of a low-resolution ADC,
we use the same architecture and $\text{FoM}$ for higher a quatization of $q=10$ bits for better comparison.

For the VGA, we consider the $90 \text{nm}$ CMOS reported in \cite{wang2012}.
The VGA's $FoM$ is 5280 for an active area of $\text{0.01mm}^2$. 
Last, we have assumed an oscillator power draw of $10 \text{~dBm}$,
a mixer insertion loss of $IL_{\rm mix}$ of $6 \text{~dBm}$ and
baseband power $P_{\rm BB}^{\rm out}$ of $10 \text{~dBm}$.

Using the above device properties, we, furthermore,
consider an operating bandwidth $(BW)$ of $400~\text{MHz}$,
the maximum provisioned bandwidth by 3GPP NR and a
maximum distance $d_{\rm max}$ of $100$ meters.
We then use the channel model \cite{AkdenizCapacity:14} to calculate the pathloss needed for computing
the received power $P_{\rm RX}$ in eq. \eqref{eq:vga_gain} and \eqref{eq:vga_gig}.
Specifically, for an EIRP Tx power of $43 \text{~dBm}$ the received power is $-87 \text{~dBm}$ at $d_{\rm max}$
where the channel is expected to be non-line-of-sight (NLOS). 
Finally, we assume a sampling rate, $F_{\rm s} = 1$~GHz and a receiver array size of $N_{\rm Rx} = 16$. While we assume a system bandwidth of $400~{\rm MHz}$, however, according to the NR specifications OFDM sampling rate is set at $491~{\rm MHz}$. Furthermore, we consider an oversampling rate at $2 \times$ the system bandwidth to avoid aliasing. Hence, the $1~{\rm GHz}$ sampling rate assumption.

By plugging the above numbers into \eqref{eq:lna_analog} -- \eqref{eq:hy_pow},
we obtain the front-end power consumption presented in Table \ref{tab:1}. 
With the term RFFE we denote the Radio Frequency front-end which is the pre-VGA part of the mmWave front-end, i.e., the LNAs and the LO. 
We observe a couple of points:
\begin{itemize}
  \item[$\circ$] The power consumption of a high-resolution
digital front-end is almost six times the consumption of
the analog front-end. 
However, it is below a factor of $N_{\rm RX}$.
This is because while there are $N_{\rm RX}$ ADCs in a digital front-end,
there are also $N_{\rm RX}$ phase shifters in an analog one,
the insertion loss of which must be compensated by the LNAs.
We will see in the next section how the gap between analog and digital
closes when we take into account the beam discovery delay.

\item [$\circ$] Reducing the ADC resolution from $q = 10$ to $q = 4$ bits
has a dramatic effect on the power consumption.
It is in fact even below that of the analog front-end. 
However, this reduction in resolution adds a distortion to the signal which
will reduce the effective SNR. This reduction has a small effect in the low
to median SNR regimes. \cite{dutta17Asilomar, dutta2019twc}.
\end{itemize}

\section{Evaluation and simulation}
\label{sec:evsim}
We now evaluate the performance of our correlation-based detector
for receivers performing analog, fully digital and low-resolution fully digital
beamforming. In a series of steps, we first define the simulation setup and parameters,
assess the effect of lowering the quantization resolution in fully digital front-ends,
and finally illustrate the interplay between the beamspace and
the energy consumption of each the front-end architectures presented in Section \ref{sec:sigsysmod}.

\subsection{Channel model and user SNR distribution}
\label{sec:chanmod}

\begin{table*}
\begin{center}
\caption{SNR distribution parameters.}
\label{tab:snr_par}
\begin{tabular}{lcl}

\hline
 {\bf Parameter} & {\bf Value} & {\bf Description}
 \tabularnewline \hline

$d_r$ & $100$~m & Cell radius \tabularnewline

$P_{TX}$ & $30$~dBm & gNB TX power \tabularnewline

$NF_{UE}$ & $7$~dB & user equipment noise figure \tabularnewline

$kT$ & $-174$~dBm/Hz & Thermal noise power density \tabularnewline

$\text{F}_c$ & 28 GHz & Carrier Frequency \tabularnewline

$W_{sys}$ & 400 MHz & System bandwidth \tabularnewline

$P_{LOS}(d)$ & $\exp (-a_{los} d)$, $ a_{los}= 67.1$~m & Probability of LOS vs. NLOS \tabularnewline

$P_{NLOS}(d)$ & $1 - P_{LOS}(d)$ & Probability of NLOS \tabularnewline

$PL(d)$ & $\mu + 10 \nu \log10(d) +\zeta, \zeta \sim \mathcal{N}(0, \xi^2)$ & Path loss model dB, $d$ in meters \tabularnewline

$\mu_{LOS}$, $\nu_{LOS}$, $\xi_{LOS}$ & $\mu=61.4$, $\nu=2.0$, $\xi^2=5.8$ dB & LOS parameters \tabularnewline

$\mu_{NLOS}$, $\nu_{NLOS}$, $\xi_{NLOS}$ & $\mu=72.0$, $\nu=2.92$, $\xi^2=8.7$ dB & NLOS parameters \tabularnewline
\hline
\end{tabular}
\end{center}
\end{table*}

We consider a single cell of radius $d_r = 100~\text{m}$, with the
mmWave base-station, or gNB, situated at the center. 
We then randomly drop user equipments in this cell and compute their pathloss to the gNB
according to the widely used model in \cite{AkdenizCapacity:14}.
This urban multipath model is based on real measurements in $28~\text{GHz}$ bands performed in 
New York City \cite{Rappaport:12-28G, McCartRapICC15, Rappaport:28NYCPenetrationLoss, Samimi:AoAD}.
While our detector was derived for a single-path LOS channel perfectly aligned
with one of the beamspace directions, we will use this multipath channel
in our evaluation and simulations. 
According to this model, links between the user equipment and the gNB are in LOS or NLOS based on
an exponential probability distribution parametrized
by the distance separating the two, $d$. 
Close user equipments have a high probability of being in LOS while as the distance grows the probability of 
being in NLOS increases.
The resulting omni-directional pathloss, i.e., before applying beamforming, in $\text{dB}$ is computed as

\beq
\label{eq:pathloss}
P_{L} = \mu + 10 \nu \log_{10}(d) + \zeta [\text{dB}], \quad \zeta \sim \mathcal{N}(0, \xi^2),
\eeq

where $\mu$, $\nu$ and $\xi^2$ are parameters defined by whether the link is LOS or NLOS. 

We then use this pathloss to derive the user equipment omni-directional SNR, $\text{SNR}_\text{omni}$ as

\beq
\label{eq:snr_omni}
\text{SNR}_\text{omni} = \frac{P_{RX}}{N_0 W_{sys}},
\eeq

where $P_{RX}$ is the received omni directional power resulted from subtracting the pathloss
in \eqref{eq:pathloss} from the transmitted power $P_{TX}$.
The remaining parameters above are the noise power density plus noise figure, $N_0$, and the system bandwidth $W_{sys}$.

\begin{figure}
\begin{center}

 \includegraphics[width=0.95\linewidth]{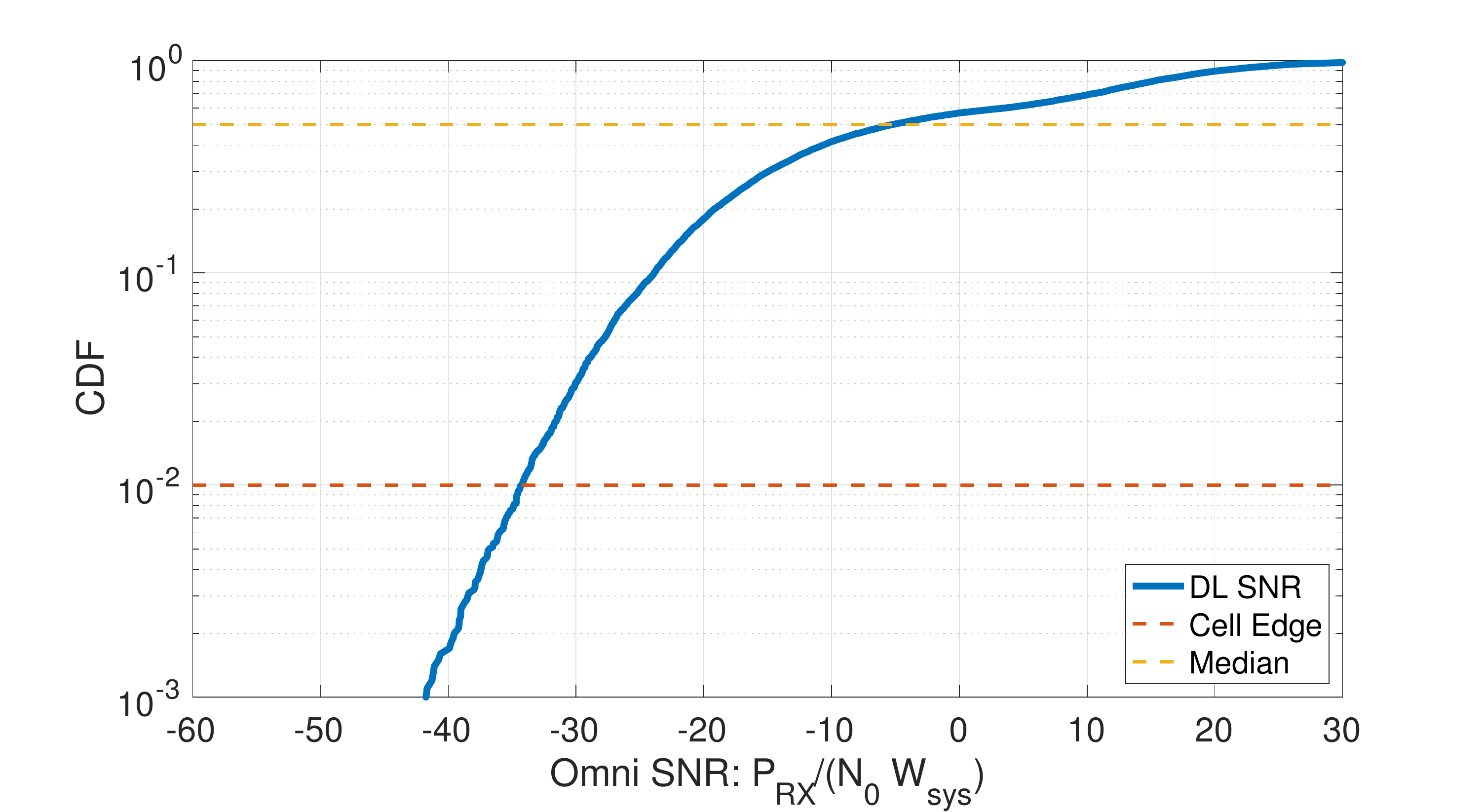}
\end{center}
\caption{SNR distribution of $10,000$ user equipment drops}
\label{fig:snr_dist}    
\end{figure}

Next, we derive the user equipments SNR distribution through $10,000$ random drops in our cell
and divide them into three regimes: cell edge user equipments, i.e., the first percentile of the CDF, 
median users and above, and those in between these regimes.
However, in the sequel, we characterize the energy consumed by the analog and digital beamformers
during beam discovery for the cell edge and the median users.
Fig. \ref{fig:snr_dist} shows the CDF of the SNR distribution.
A summary of the parameters and their values used to derive this distribution
is presented in Table \ref{tab:snr_par}. 

\subsection{Low-resolution fully digital beamforming }
\label{sec:low_res_dig}
In section \ref{sec:rf_arch}, we presented the power consumption of four mmWave front-ends
implementing analog, hybrid and fully digital beamforming . 
We also showed that the power consumption of the fully digital front-end can be dramatically reduced by
employing low-resolution ADCs. 

However, reducing the resolution will degrade the quality of the received signal.
We quantify this degradation by using the method presented in \cite{barati2015directional, ChenG:87, FletcherRGR:07}.
According to \cite{FletcherRGR:07}, the effect of low-resolution quantization, i.e., analog to digital conversion, is
modeled as a reverse gain $(1-\gamma)$ multiplied with the received sample if it had gone through an
infinite-resolution quantizer and an additive Gaussian noise term.
That is if the high quantization sample is given as

\[
y[n] = x[n] + w[n],
\]
where$x[n]$ are the transmitted signal sample and $w[n]$ AWG noise with some variance $\Exp|{w[n]|^2 = \sigma_n^2}$,
then the imperfectly quantized received sample is given as

\beq
\label{eq:y_quant}
y_q[n] = (1 - \gamma)x[n] + (1 - \gamma) w[n] + v[n]. 
\eeq

The last term $v[n]$ is a zero-mean Gauusian random variable with the following variance \cite{barati2015directional, FletcherRGR:07}

\beq
\label{eq:q_var}
\sigma_{\rm q}^2 =\Exp{|v [n]|^2} = \gamma (1-\gamma) ( \Exp{|x[n]|^2} + \sigma_n^2).
\eeq

Now, using eq. \eqref{eq:y_quant} the effective SNR ${\rm SNR}_{\rm eff}$ after quantization is obtained as

\[
{\rm SNR}_{\rm eff} = \frac{(1 - \gamma ) {\rm SNR_{\infty}}}{ 1+ \gamma {\rm SNR_{\infty}}}, 
\]
where ${\rm SNR_{\infty}}$ is the SNR value of an infinite-resolution quantizer. 
The value of $\gamma$ depends on the quantizer's design, the input distribution and the
quantization resolution bits $q$.
As an example, for a Gaussian input the value of $\gamma$ is only $0.01$. 

Thus, we see that low-quantization fully
digital beamforming comes with a penalty in the SNR, an observation corroborated by the theoretic work in \cite{MoHeathTsp15} and \cite{Singh2009Limits}.
However, this penalty can be negligible depending on the quantization bits $q$, and the SNR regime,
as shown in \cite{dutta17Asilomar} and \cite{dutta2019twc}, where the same model was used for OFDM inputs.
The effective SNR can be much lower, though, for very high values of ${\rm SNR_{\infty}}$ at very low quantization bits, say one or two. 
However, in cellular systems, the overwhelming majority of user equipments are at low to medium SNR regimes
where the distance between the effective SNR and ${\rm SNR_{\infty}}$ is below $3~{\rm dB}$.
Furthermore, at a quantization of $q = 4$ bits as we consider for our low-resolution fully digital front-end,
even at ${\rm SNR_{\infty}}$ as high as $20~{\rm dB}$ the effective SNR is still below $3 ~{\rm dB}$ \cite{dutta17Asilomar, dutta2019twc}.

\subsection{System parameters}
\label{sec:sysparam}

\begin{table*}
\begin{center}
\caption{System parameters.}
\label{tab:sys_par}
\begin{tabular}{ll}
\hline
 {\bf Parameter} & {\bf Value}
 \tabularnewline \hline

Total system bandwidth, $W_{sys}$ & $400$~MHz
\tabularnewline

Signal Duration, $T_{\rm PSS}$ & $8.91 {\rm \mu s}$ ($1$ OFDM symbol)
\tabularnewline

Sub-carrier spacing, $\Delta_{\rm f}$ & $120 {\rm kHz}$
\tabularnewline

PSS Bandwidth, $W_{\rm PSS}$ & $15.24 {\rm MHz}$ (127 sub-carriers)
\tabularnewline

SS burst period, $T$ & $20 {\rm ms}$
\tabularnewline 

SS burst duration, $T_{\rm ssb}$ & $5 {\rm ms}$
\tabularnewline 

Total false alarm rate per scan cycle,
$R_{\textsc FA}$ & $ 0.01$
\tabularnewline

Number of PSS waveform hypotheses $N_{\rm PSS}$ & $3$
\tabularnewline

Number of frequency offset hypotheses, $N_{\textsc FO}$ & $ 3 $ \tabularnewline 

gNB antenna & $N_{\rm TX} \times N_{\rm TX} $ uniform planar array  \tabularnewline 
UE antenna & $ N_{\rm RX} \times N_{\rm RX}$ uniform planar array  \tabularnewline 

gNB antenna elements in each dimension & $N_{rm TX} = 8$  \tabularnewline 
UE antenna elements in the azimuth dimension & varies: $N_{\rm RX}^{\rm az} = 2, 4$  \tabularnewline 
UE antenna elements in the elevation dimension & varies: $N_{\rm RX}^{\rm el} = 2, 4$  \tabularnewline 

gNB directions & $N_D^{TX} = N_{\rm TX} \times N_{\rm TX} = 8 \times 8 = 64$  \tabularnewline 
UE directions & varies $N_D^{RX} = N_{\rm RX}^{\rm az} \times N_{\rm RX}^{\rm el}= 4, 8, 16$   \tabularnewline 
gNB beamforming & Hybrid $M = 2$ \tabularnewline 
UE beamforming & Varied: Analog, Digital, low-res. Digital  \tabularnewline \hline

\end{tabular}
\end{center}
\end{table*}

We will compare the mmWave front-ends presented in Section \ref{sec:sigsysmod}, within the context 
of 3GPP NR's PSS discovery. 
We assume a PSS signal that is confined within one OFDM symbol and 127 sub-carriers.
For a sub-carrier spacing of $\Delta_f = 120 {\rm kHz}$, they are $8.91 {\rm \mu s}$
and 15.24 ${\rm MHz}$, respectively. 
We will assume that the channel is almost flat within the duration and bandwidth of the signal.

Both the user equipment and the gNB are equipped with uniform planar arrays that allow them to beamsteer in both azimuth (az) and elevation (el).
Note that from now on we the note the available directions at the gNB and user equipment with
$N_D^{\rm Tx}$ and $N_D^{\rm Rx}$, respectively.
These are the products of $N_D^{\rm Tx}$ and $N_D^{\rm Rx}$ in each dimension of the 2D arrays. 
Hence, the beamspace size $L$ is equal to $N_D^{\rm Tx} \times N_D^{\rm Rx}$.
The gNB is assumed to perform only hybrid beamforming with $M = 2$ directions at a time,
while the user equipment may employ one of analog, fully digital and low-resolution fully digital beamforming .
Note that we keep the transmission power of the gNb fixed at $30~{\rm ~dBm}$ (Table \ref{tab:snr_par}).
Therefore, transmitting the PSS in $M = 2$ directions implies that the sum of
the power transmitted in each direction must be $30~{\rm ~dBm}$.
We will assume equal power transmission. Thus, the signal in each direction
will carry half of the total $30~{\rm ~dBm}$ power.
The resulting beamforming gains will be applied to the user equipments' omni-directional SNR given by eq. \eqref{eq:snr_omni}
and the effective SNR in case of the low-resolution digital front-end. 

Also, we assume a beamspace discovery process as described in Section \ref{sec:nr_beamdis},
where the gnB transmits the PSS directionally in its $N_D^{\rm Tx}$ direction within
an SS burst of duration $5 {\rm ms}$ every $T = 20~{\rm ms}$.
The user equipment also sweeps its $N_D^{\rm Rx}$ directions in search of one of the $N_{\rm PSS} = 3$ 
PSS sequences. 
Finally we compute the threshold $\theta$ in test \eqref{eq:log_test}.
We will do so through the false alarm probability $P_{\rm FA}$.
Suppose we would like to maintain a constant target false alarm rate of $R_{\rm FA}$ during each searching period.
Then, the false alarm probability and the false alarm rate are connected through 

\beq
\label{sec:pfa}
P_{\rm FA} = \frac{R_{\rm FA}}{N_{\rm PSS} N_{\rm dly} N_{\rm FO}}.
\eeq

Now, $N_{\rm dly}$ is the number of delay hypotheses. 
That is at which delay $tau$ the signal was present. 
Since the PSS transmission is periodic, the number of the delays hypotheses is bounded
and it is the number of the samples within each SS burst period.
If the signal was present at a delay $\tau_1$, then it is also present at
delay $T + \tau_1$, where $T$ is the SS burst period. 

To compensate for two other sources of uncertainty, we incorporate them into the
$P_{FA}$ calculation. These are the number of possible PSS sequences $N_{\rm PSS}$ and the
number of frequency offsets hypotheses $N_{FO}$.
There are $N_{\rm PSS}$ sequences that the user equipment may
mistake one with the other two and are factored in the false alarm calculation.
Next,we approximate $N_{FO}$ as follows.
First, we assume a local oscillator error of $\pm 10$ parts per million (ppm) and
a Doppler shift for a velocity up to $30~{\rm km/h}$ at $F_c = 28~{\rm GHz}$. 
The LO error and the Doppler shift will define a range $\pm \Delta_{fmax}$. 
We then discretize this range so that in each interval the channel does not rotate more than $\pi/4$ within the duration of one OFDM symbol.
The number of frequency offset hypotheses is then 
the number of obtained intervals. 

The threshold $\theta$ therefore is the value of the normalized correlation
$\rho_{k\ell}$ which corresponds to the target $P_{FA}$. 
The detector will consider everything above this threshold as received signal and anything below as noise.

All the parameters described above are tabulated in Table \ref{tab:sys_par}.

\begin{figure}
\begin{center}

 \includegraphics[width=0.95\linewidth]{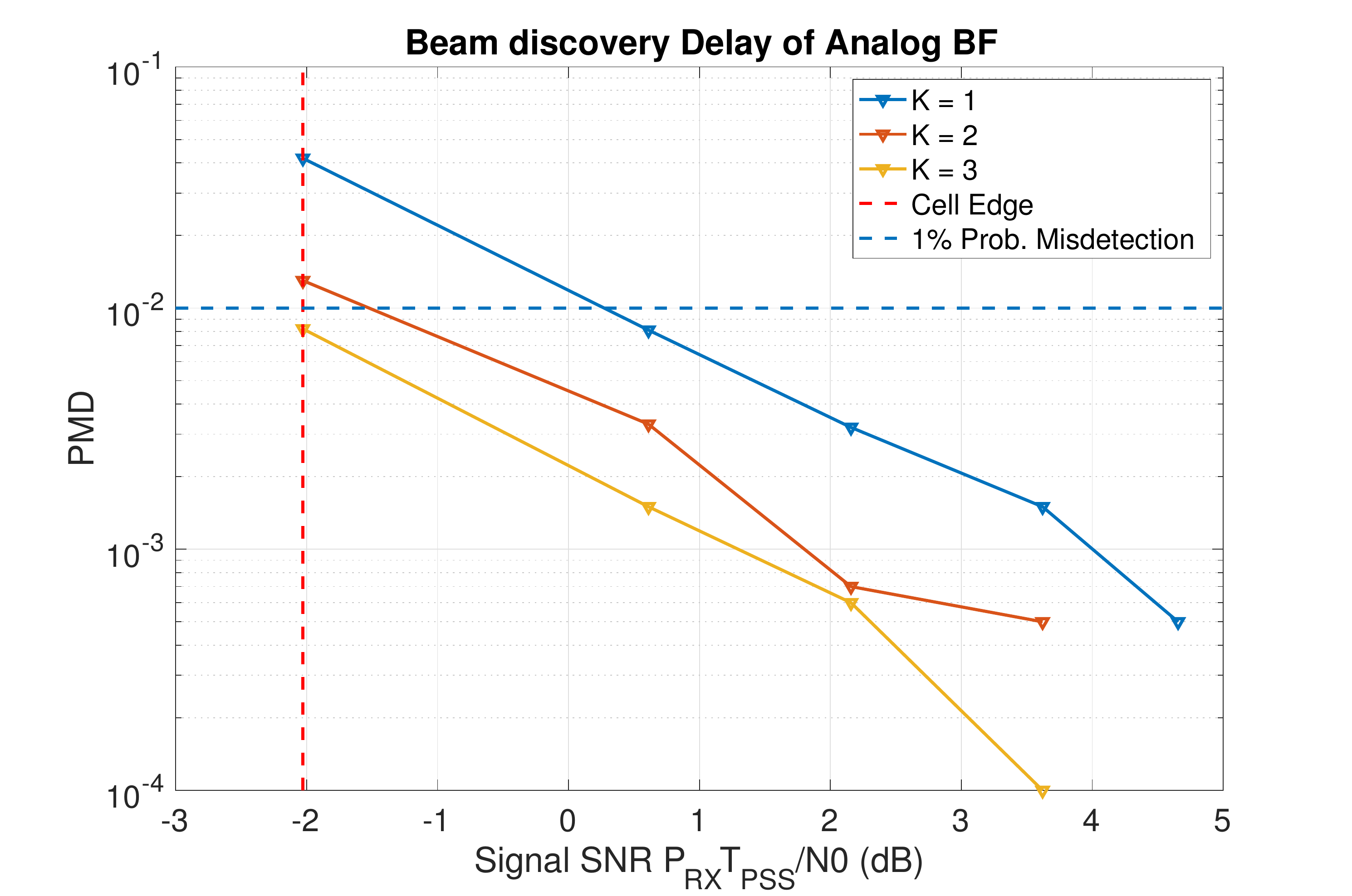}
\end{center}
\caption{Analog Receiver: Mis-detection vs Signal SNR. $K = 1,2,3$ and $P_{\rm MD}^{\rm tgt} = 1\%$.}
\label{fig:pmd}    
\end{figure}

\subsection{Comparison of different beamforming architectures}
\label{sec:bf_comp}

\begin{table}
\begin{center}
\caption{Upper bounds of initial discovery delays for different mmWave front-ends. $P_{\rm MD}^{\rm tgt}$ is set to $1\%$.}
\label{tab:pmd_delay}    
\begin{tabular}{lcc}
\hline\noalign{\smallskip}
{\bf front-end Arch.} & {\bf Cell Edge Disc. Delay} & {\bf Median User Disc. Delay}  \\
\noalign{\smallskip}\hline\noalign{\smallskip}
Analog       & $960~{\rm ms}$ & $320~{\rm ms}$ \\
Digital (High res.)& $60~{\rm ms}$ & $20~{\rm ms}$ \\
Digital (Low res.) & $80~{\rm ms}$ & $20~{\rm ms}$ \\

\noalign{\smallskip}\hline
\end{tabular}
\end{center}
\end{table}

Now that we have defined the channel model, system and signal parameters, and
described the beamspace and bean discovery procedure, we can start the comparison
between the mmWave front-ends presented in Section \ref{sec:rf_arch}.
We first look into the beam discovery delays with beamsweeping for the beamformers and then tie
these delays to the power consumptions reported in Section \ref{sec:fr_powcons}. 

\subsubsection{Beamsweeping delay}
\label{sec:del_comp}

We estimate the initial discovery delay as the time needed to go through all the beams in the beamspace 
times the total number beamsweeps $K$ necessary to determine the best direction for a target mis-detection
probability $P_{\rm MD}^{\rm tgt}$.
Hence, $K$ is a function of $P_{\rm MD}^{\rm tgt}$.
Since our detector is essentially an energy detector, maximizing energy of the incoming signal from the correct
direction, depending on the SNR regime the user equipment may need to perform multiple sweeps. 
In Fig. \ref{fig:pmd} we show the number of beamsweeps necessary to discover
the correct path to the gNB for a $P_{\rm MD}^{\rm tgt} = 1\%$. 
A mobile user at the cell edge needs $K = 3$ beamsweeps to get the direction right,
while the rest of the users can detect the signal in one beamsweep. 

Take as an example the case of the analog receiver with $N_D^{\rm Rx} = 16$ directions.
The size $L$ of the beamspace is then $16 \times 64 = 1024$.
Then, the initial discovery delay $T_{\rm delay}(P_{\rm MD}^{\rm tgt})$, for such a user equipment to successfully discover the true direction is

\beqa
\label{eq:dly_an}
\frac{(K(P_{\rm MD}^{\rm tgt}) - 1 )\times L \times T}{N_{\rm D}^{\rm Tx}} + T_{\rm ssb}  \leq T_{\rm delay}(P_{\rm MD}^{\rm tgt}) \leq \nonumber \\ \leq \frac{K(P_{\rm MD}^{\rm tgt})\times L \times T}{N_{\rm D}^{\rm Tx}} \nonumber \\
= \frac{20480 K(P_{\rm MD}^{\rm tgt}) }{64} {\rm ms},
\eeqa
where $T = 20~{\rm ms}$ is the SS burst period and $T_{\rm ssb} = 5~{\rm ms}$ the duration of an SS burst.
The reason we divided the quantity above with $N_D^{\rm Tx}$ is that within each SS burst period $T$,
the gNB sweeps all its directions in one SS burst.
Also, notice that we did not divide $L$ by the number of the RF chains $M =2$ at the transmitter.
Since the NR procedure defines a maximum of $B= 32$ SS blocks within one SS burst and the number of
directions the gNB scans is $N_{\rm D}^{\rm Tx} = 64$, the reduction in the effective beamspace
due to hybrid beamforming is accounted for.
Otherwise, dividing eq. \eqref{eq:dly_an} by $M$ would imply that the user equipment probes $M = 2$ directions
directions each time, which is not correct since it employs an analog beamformer.

From eq. \eqref{eq:dly_an}, we can see that even for a user that is not at the cell edge, i.e., 
$K (P_{\rm MD}^{\rm tgt}) = 1$, the initial discovery delay can go as high as $320 {\rm ms}$,
when the entire control-plane delay of 4G LTE must be below $50{\rm ms}$ \cite{3GPP36913}.
For the cell edge user on the other hand, this delay is between $640 ~{\rm ms}$ and almost a second.
This highlights the large impact of the beamspace size on the initial discovery delay. 
Now, during initial discovery the user equipment must be constantly ``on'' searching for the correct beamspace.
This amount of delay as we will show next has a large impact on the power consumption of the analog beamformer. 

In Table \ref{tab:pmd_delay} following the same process and using eq. \eqref{eq:dly_an},
we give the upper bounds of the initial discovery delay for the
other two kinds of front-ends, fully digital and low resolution fully digital.
All the numbers are for a user equipment array size of $16$ elements.
Notice that the SNR penalty due to low quantization is negligible in both the cell edge and the rest of the area. 
The large gain compared to analog was the result of shrinking the beamspace size
$L$ to only $N_{\rm D}^{\rm Tx}$ which is all covered in one SS burst.

\subsubsection{Beamsweeping Energy consumption}
\label{sec:enrerg_comp}

\begin{figure}
\begin{center}
 \includegraphics[width=.5\textwidth]{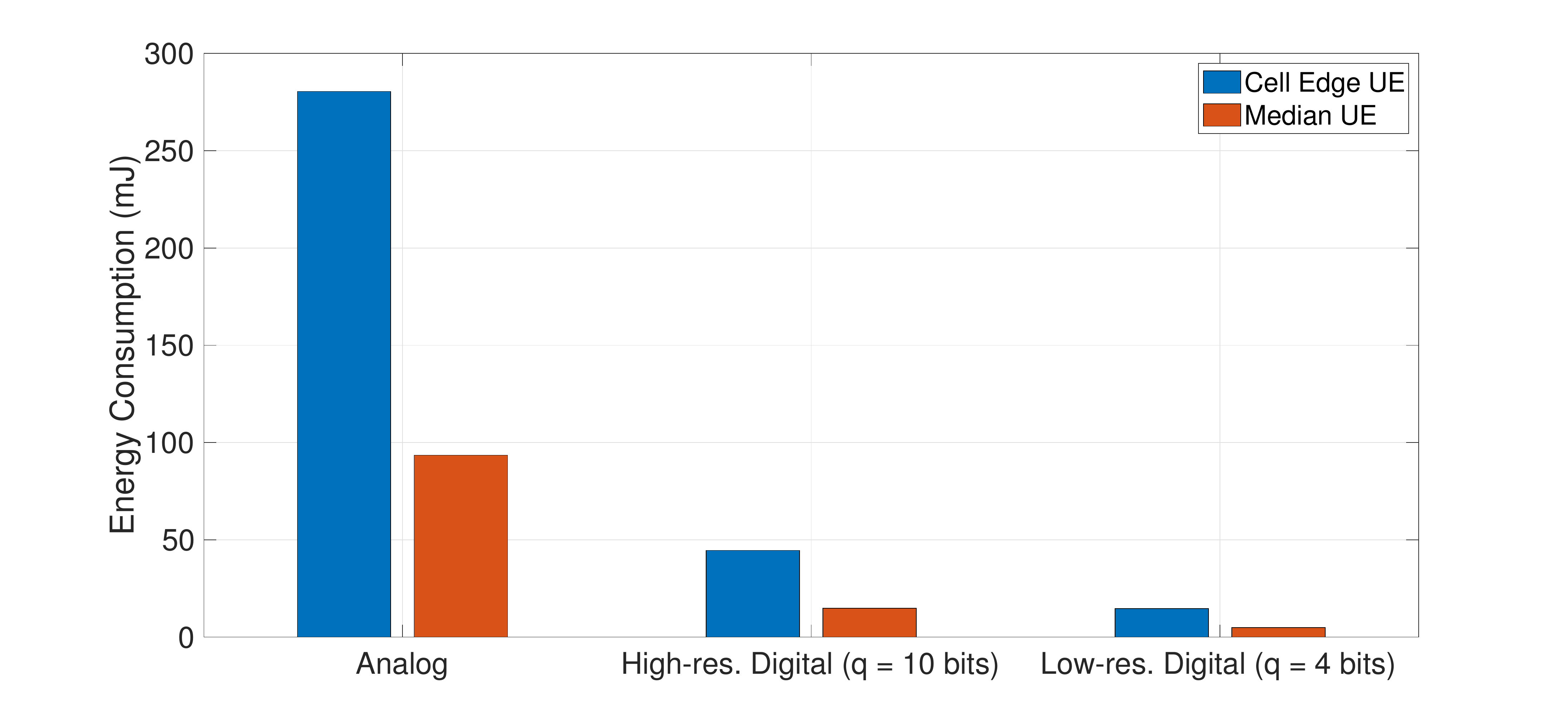}
\end{center}
\caption{Energy consumption of the mmWave front-ends: Analog, High-resolution Digital and Low-resolution Digital. Number of antenna elements at the user equipment $N_{\rm Rx} = 16$.
Cell Edge user equipment in blue, Median user equipment in red.}
\label{fig:en_bar}    
\end{figure}

\begin{figure*}[!t]
\centering
\subfloat [Cell Edge user equipment] {
	\includegraphics[width=0.45\textwidth]{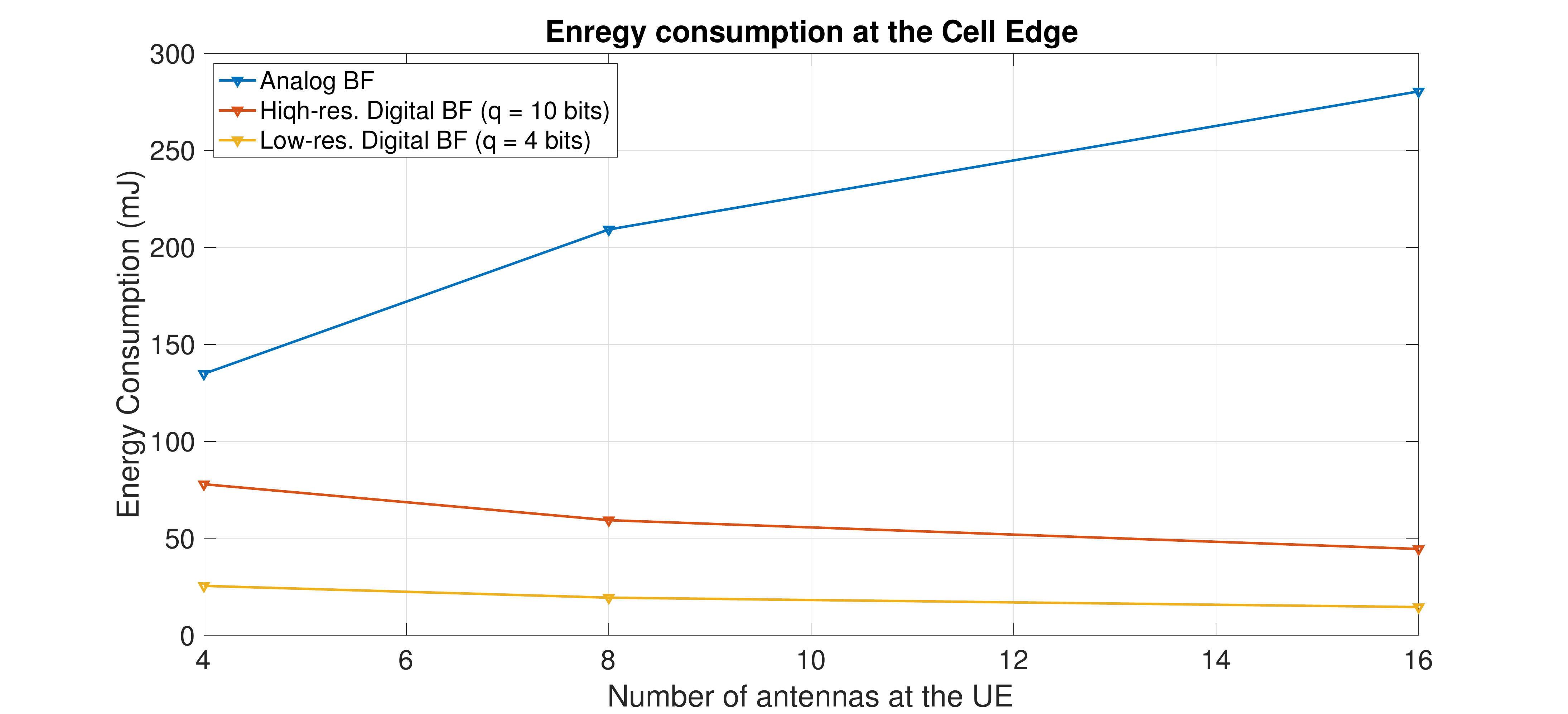} \label{fig:nant_edge}
} %
\subfloat [Median user equipment] {
	\includegraphics[width=0.45\textwidth]{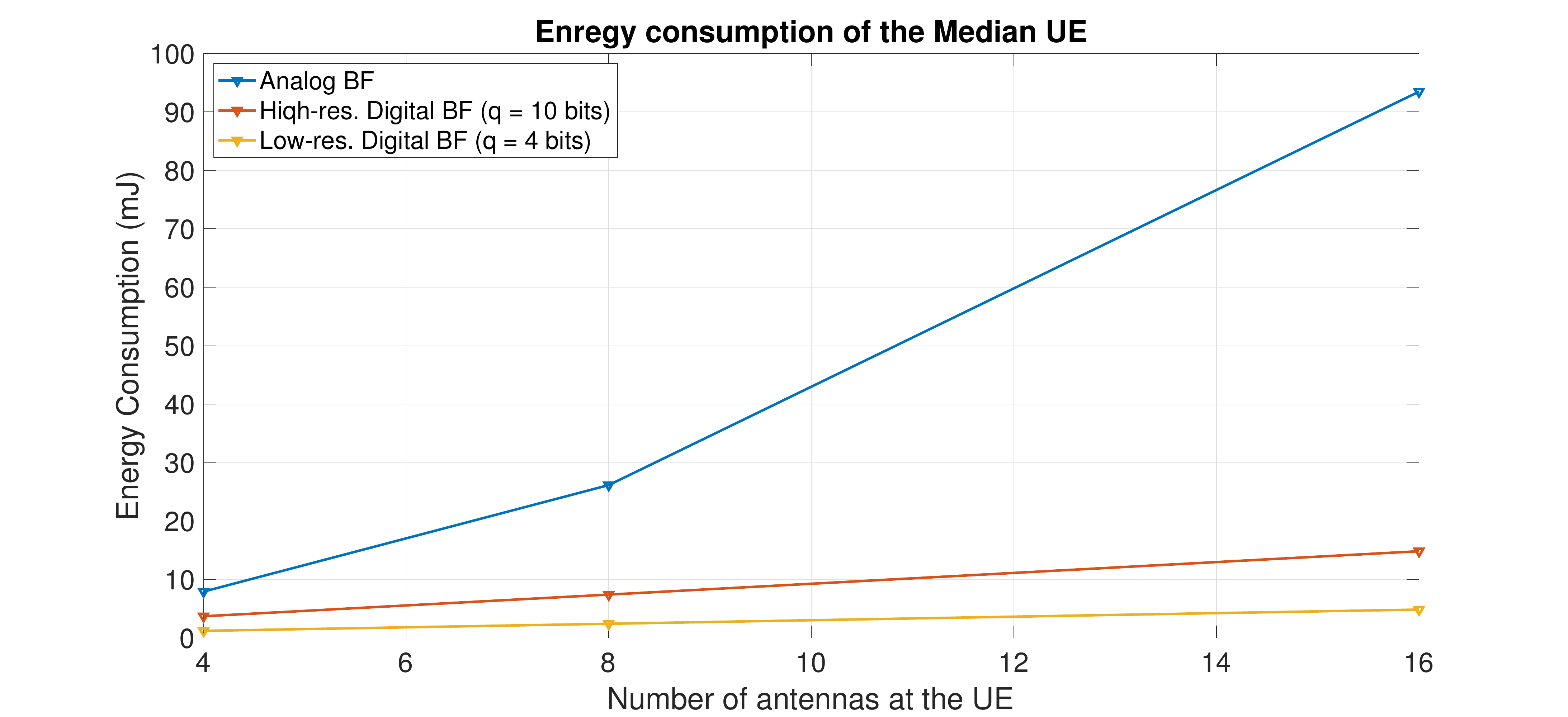} \label{fig:nant_med}
}
\caption{Energy consumption as a function of the user equipment aray size. Fig. (\ref{fig:nant_edge}) Energy consumption at the cell edge. Fig. (\ref{fig:nant_med}) energy consumption of the median user equipment.}
\label{fig:nant_eng}
\end{figure*}

We have so far characterized the beam discovery delay of the analog,
the high-resolution digital and the low-resolution digital beamformers.
The energy consumption of a front-end is tightly connected to the amount of time
it needs to operate. 
Since the user equipment is assumed to be always ``on'' trying to attach to a gNB,
it is necessary to go beyond the device power consumption and look at energy consumption
during the time needed to establish a directional link between the user equipment and the gNB. 
In Fig. \ref{fig:en_bar} we show the energy consumption of the three front-ends,
where the user equipment array size is $ N_{\rm RX} = 16$. 
The difference between analog and the two versions of digital is astonishing.
The energy consumption of analog towers even that
of high-resolution digital.
Remember, when we looked at the front-ends' power consumption (Jouls/s) in Section \ref{sec:fr_powcons}
digital beamforming 's consumption was six times that of analog beamforming .
What Fig. \ref{fig:en_bar} shows though, is the
analog beamforming front-end burning almost four times
more energy than the high-resolution digital front
end during beamsweeping.

However, since after establishing the directional link, in the data communication phase,
analog is expected to burn less power than fully digital, we believe that mmWave systems will need
to use a low-resolution fully digital front-end.
Despite the fact that the difference in energy consumption between high and low-resolution digital front-ends
is quite small during beam discovery.

Next, we turn our attention to the number of antenna elements at the receiver. 
We would like to know how the energy consumption scales with the size of the antenna
array in the initial beamsweeping phase.
In other words, what is the impact of various beamspace sizes on the energy consumption?
Fig. \ref{fig:nant_eng} answers this question.
We try three different configurations at the user equipment: a 2D array of size 4, a 2D array of size 8,
and a 2D array of size 16. 
There are a few interesting points these two figures raise:
\begin{itemize}
  \item [$\circ$] First, while fewer antennas do bring the energy consumption of
  the user equipment down, analog beamforming still burns considerably more energy than both the digital front-ends.
  \item[$\circ$] Second, the gap between analog and digital beamforming widens as more antennas are employed at the user equipment.
  The reason is the following: when the number of antennas is small, and
  the post-BF SNR is low, analog beamforming 
  looks into a fewer directions at each period. This means that the most
  crucial factor in the front-ends' energy consumption is
  the size of the beamspace. This leads to higher beam discovery delays
  resulting in longer ``on'' time for the user equipment.
  This is more evident in Fig. \ref{fig:nant_med} where all front-ends need
  no more than $K= 1$ beamsweep to detect the true path to the gNB.
  
  \item[$\circ$] Related to the previous point, the value of digital beamforming is evident in every antenna configuration.
  The reduction in the size of the effective beamspace is the fundamental and necessary aspect of expediting beam discovery
  and subsequently, lowering energy consumption.
  Beyond initial link establishment, digital beamforming will reduce the complexities involved in handovers and avoiding blockage in connected mode.
  
\end{itemize}

\section{Conclusion and future directions}
\label{sec:conc}
The realization of mmWave communication systems requires addressing two
crucial challenges, namely beam discovery delay and energy consumption.
In this work, we have revealed how closely these two are tied together.
Specifically, we showed that employing analog beamforming can significantly increase both the discovery delay and, subsequently,
energy consumption.
Hence, analog beamforming does not buy us lower energy consumption.

Digital beamforming, on the other hand, achieves lower delays and
lower energy consumption during beam discovery.
In addition, depending on operating SNR, using digital beamforming can allow
the use of multi-stream communication and allow joint flexible scheduling
of frequency resources and directional beams.
This may be extremely useful in the case of small data packets.
By expediting beam discovery, fully digital may enable more aggressive use of sleep mode since beam tracking and paging can be less frequent.
In fact, studying the effect of beamforming architectures on higher layers,i.e., MAC and TCP/IP, in a multi-cell setup is an interesting direction for future research.

It is true that when the directional data link is established, analog burns less energy compared to fully digital.
Reducing the quantization resolution of the ADC pairs, however, can bring
the energy requirements in data communication at the level of analog beamforming 
at a penalty negligible in most SNR regimes.
Thus, low-resolution digital beamforming enables the realization of mmWave cellular
systems with low discovery delays while allowing
more flexible scheduling. 

%
%


%
%


\bibliographystyle{spmpsci}   
\bibliography{bibl}  

\newcommand{\SortNoop}[1]{}
\begin{thebibliography}{10}
\providecommand{\url}[1]{{#1}}
\providecommand{\urlprefix}{URL }
\expandafter\ifx\csname urlstyle\endcsname\relax
  \providecommand{\doi}[1]{DOI~\discretionary{}{}{}#1}\else
  \providecommand{\doi}{DOI~\discretionary{}{}{}\begingroup
  \urlstyle{rm}\Url}\fi

\bibitem{3GPP36913}
3GPP: {Requirements for Further Advancements for E-UTRA (LTE-Advanced)}.
\newblock TR 36.913 (release 9) (2010)

\bibitem{3GPP_NRphyMod}
3GPP: {NR} -- {P}hysical {C}hannels and {M}odulation.
\newblock TS 38.211 (release 15) (2018)

\bibitem{3GPP_NRphyPhyMeas}
3GPP: {NR}--{P}hysical {L}ayer {M}easurements.
\newblock TS 38.215 (release 15) (2018)

\bibitem{3GPP_NRphyProcCont}
3GPP: {NR}--{P}hysical {L}ayer {P}rocedures for {C}ontrol.
\newblock TS 38.213 (release 15) (2018)

\bibitem{3GPP_NRphyRRC}
3GPP: {NR}--{R}adio {R}esource {C}ontrol ({RRC}) {P}rotocol {S}pecification.
\newblock TS 38.331 (release 15) (2018)

\bibitem{abbas_context2016}
{Abbas}, W.B., {Zorzi}, M.: Context information based initial cell search for
  millimeter wave 5g cellular networks.
\newblock In: 2016 European Conference on Networks and Communications (EuCNC),
  pp. 111--116 (2016)

\bibitem{Adabi07}
Adabi, E., Heydari, B., Bohsali, M., Niknejad, A.M.: 30 {GHz CMOS} low noise
  amplifier.
\newblock In: Proc. IEEE RFIC Symp., pp. 625--628 (2007)

\bibitem{AkdenizCapacity:14}
Akdeniz, M., Liu, Y., Samimi, M., Sun, S., Rangan, S., Rappaport, T., Erkip,
  E.: Millimeter wave channel modeling and cellular capacity evaluation.
\newblock IEEE J. Sel. Areas Commun. \textbf{32}(6), 1164--1179 (2014)

\bibitem{alkhateeb_init2017}
{Alkhateeb}, A., {Nam}, Y., {Rahman}, M.S., {Zhang}, J., {Heath}, R.W.: Initial
  beam association in millimeter wave cellular systems: Analysis and design
  insights.
\newblock IEEE Trans. on Wireless Commun. \textbf{16}(5), 2807--2821 (2017)

\bibitem{Rappaport:12-28G}
Azar, Y., Wong, G.N., Wang, K., Mayzus, R., Schulz, J.K., Zhao, H., Gutierrez,
  F., Hwang, D., Rappaport, T.S.: 28 {GHz} propagation measurements for outdoor
  cellular communications using steerable beam antennas in {N}ew {Y}ork {C}ity.
\newblock In: Proc.\ IEEE ICC (2013)

\bibitem{barati2015directional}
Barati, C.N., Hosseini, S., Rangan, S., Liu, P., Korakis, T., Panwar, S.,
  Rappaport, T.S.: Directional cell discovery in millimeter wave cellular
  networks.
\newblock IEEE Trans. Wireless Commun. \textbf{14}(12), 6664 -- 6678 (2015)

\bibitem{barati2016initial}
Barati, C.N., Hosseini, S.A., Mezzavilla, M., Korakis, T., Panwar, S.S.,
  Rangan, S., Zorzi, M.: Initial access in millimeter wave cellular systems.
\newblock IEEE Trans.\ Wireless Commun \textbf{15}(12), 7926--7940 (2016)

\bibitem{booth_bandit2019}
{Booth}, M.B., {Suresh}, V., {Michelusi}, N., {Love}, D.J.: Multi-armed bandit
  beam alignment and tracking for mobile millimeter wave communications.
\newblock IEEE Communications Letters \textbf{23}(7), 1244--1248 (2019)

\bibitem{capone_cont2015}
{Capone}, A., {Filippini}, I., {Sciancalepore}, V.: Context information for
  fast cell discovery in mm-wave 5g networks.
\newblock In: Proceedings of European Wireless 2015; 21th European Wireless
  Conference (2015)

\bibitem{capone_obst2015}
{Capone}, A., {Filippini}, I., {Sciancalepore}, V., {Tremolada}, D.: Obstacle
  avoidance cell discovery using mm-waves directive antennas in 5g networks.
\newblock In: 2015 IEEE 26th Annual International Symposium on Personal,
  Indoor, and Mobile Radio Communications (PIMRC), pp. 2349--2353 (2015)

\bibitem{ChenG:87}
Chen, J.H., Gersho, A.: Gain-adaptive vector quantization with application to
  speech coding.
\newblock IEEE Trans. Commun. \textbf{COM-35}(9), 918--930 (1987)

\bibitem{chiu_hier2019}
{Chiu}, S., {Ronquillo}, N., {Javidi}, T.: Active learning and csi acquisition
  for mmwave initial alignment.
\newblock IEEE J. Sel. Areas Commun. \textbf{37}(11), 2474--2489 (2019)

\bibitem{dedonno_2017}
{De Donno}, D., {Palacios}, J., {Widmer}, J.: Millimeter-wave beam training
  acceleration through low-complexity hybrid transceivers.
\newblock IEEE Transactions on Wireless Communications \textbf{16}(6),
  3646--3660 (2017)

\bibitem{devoti_cont2018}
{Devoti}, F., {Filippini}, I., {Capone}, A.: Mm-wave initial access: A context
  information overview.
\newblock In: 2018 IEEE 19th International Symposium on "A World of Wireless,
  Mobile and Multimedia Networks" (WoWMoM), pp. 1--9 (2018)

\bibitem{dutta17Asilomar}
Dutta, S., Barati, C.N., Dhananjay, A., Rangan, S.: 5{G} millimeter wave
  cellular system capacity with fully digital beamforming.
\newblock In: Proc. Asilomar Conf. on S, S \& C, pp. 1224--1228 (2017)

\bibitem{dutta2019twc}
{Dutta}, S., {Barati}, C.N., {Ramirez}, D., {Dhananjay}, A., {Buckwalter},
  J.F., {Rangan}, S.: A case for digital beamforming at mmwave.
\newblock IEEE Trans. Wireless Commun. pp. 1--1 (2019)

\bibitem{eliasi_lowTWC}
{Eliasi}, P.A., {Rangan}, S., {Rappaport}, T.S.: Low-rank spatial channel
  estimation for millimeter wave cellular systems.
\newblock IEEE Trans. Wireless Commun. \textbf{16}(5), 2748--2759 (2017)

\bibitem{filipini_cont2018}
{Filippini}, I., {Sciancalepore}, V., {Devoti}, F., {Capone}, A.: Fast cell
  discovery in mm-wave 5g networks with context information.
\newblock IEEE Transactions on Mobile Computing \textbf{17}(7), 1538--1552
  (2018)

\bibitem{FletcherRGR:07}
Fletcher, A.K., Rangan, S., Goyal, V.K., Ramchandran, K.: Robust predictive
  quantization: Analysis and design via convex optimization.
\newblock IEEE J. Sel. Topics Signal Process. \textbf{1}(4), 618--632 (2007)

\bibitem{garcia_inband2018}
{Garcia}, G.E., {Seco-Granados}, G., {Karipidis}, E., {Wymeersch}, H.:
  Transmitter beam selection in millimeter-wave mimo with in-band
  position-aiding.
\newblock IEEE Transactions on Wireless Communications \textbf{17}(9),
  6082--6092 (2018)

\bibitem{giordani2018}
Giordani, M., Polese, M., Roy, A., Castor, D., Zorzi, M.: A tutorial on beam
  management for 3gpp nr at mmwave frequencies.
\newblock IEEE Communications Surveys Tutorials \textbf{21}(1), 173--196 (2019)

\bibitem{guo_gen2017}
{Guo}, H., {Makki}, B., {Svensson}, T.: A genetic algorithm-based beamforming
  approach for delay-constrained networks.
\newblock In: 2017 15th International Symposium on Modeling and Optimization in
  Mobile, Ad Hoc, and Wireless Networks (WiOpt), pp. 1--7 (2017)

\bibitem{hashemi_2018_cont}
Hashemi, M., Sabharwal, A., Koksal, C., Shroff, N.: Efficient beam alignment in
  millimeter wave systems using contextual bandits.
\newblock In: Proc. IEEE INFOCOM, pp. 2393--2401 (2018)

\bibitem{Hussain_code2017}
Hussain, M., Love, D.J., Michelusi, N.: Neyman-pearson codebook design for beam
  alignment in millimeter-wave networks.
\newblock In: Proceedings of the 1st ACM Workshop on Millimeter-Wave Networks
  and Sensing Systems 2017, mmNets '17, pp. 17--22. ACM (2017)

\bibitem{wifi_80211ad}
IEEE: {IEEE} standard for information technology?telecommunications and
  information exchange between systems local and metropolitan area
  networks?specific requirements - part 11: Wireless lan medium access control
  (mac) and physical layer (phy) specifications.
\newblock IEEE Std 802.11-2016 (Revision of IEEE Std 802.11-2012) pp. 1--3534
  (2016)

\bibitem{jaesim_code2019}
{Jaesim}, A., {Siasi}, N., {Aldalbahi}, A., {Ghani}, N.: Beam-bundle codebook
  for highly directional access in mmwave cellular networks.
\newblock IEEE Communications Letters \textbf{23}(11), 2104--2108 (2019)

\bibitem{McCartRapICC15}
MacCartney, G.R., Samimi, M.K., Rappaport, T.S.: Exploiting directionality for
  millimeter-wave system improvement.
\newblock In: International Conference on Communications (ICC), 2015 IEEE
  (2015)

\bibitem{marandi_game2019}
{Marandi}, M.K., {Rave}, W., {Fettweis}, G.: Beam selection based on sequential
  competition.
\newblock IEEE Signal Processing Letters \textbf{26}(3), 455--459 (2019)

\bibitem{marcano_macrocell2016}
{Marcano}, A.S., {Christiansen}, H.L.: Macro cell assisted cell discovery
  method for 5g mobile networks.
\newblock In: 2016 IEEE 83rd Vehicular Technology Conference (VTC Spring), pp.
  1--5 (2016)

\bibitem{MoHeathTsp15}
Mo, J., Heath, R.W.: Capacity analysis of one-bit quantized {MIMO} systems with
  transmitter channel state information.
\newblock IEEE Trans. Signal Process. \textbf{63}(20), 5498--5512 (2015)

\bibitem{Nasri2017}
Nasri, B., Sebastian, S.P., You, K.D., RanjithKumar, R., Shahrjerdi, D.: A 700
  {uW} 1{GS/s} 4-bit folding-flash {ADC} in 65nm {CMOS} for wideband wireless
  communications.
\newblock In: Proc. ISCAS, pp. 1--4 (2017)

\bibitem{palacios_2016}
{Palacios}, J., {De Donno}, D., {Giustiniano}, D., {Widmer}, J.: Speeding up
  mmwave beam training through low-complexity hybrid transceivers.
\newblock In: 2016 IEEE 27th Annual International Symposium on Personal,
  Indoor, and Mobile Radio Communications (PIMRC), pp. 1--7 (2016)

\bibitem{park_locbase2017}
{Park}, E., {Choi}, Y., {Han}, Y.: Location-based initial access and beam
  adaptation for millimeter wave systems.
\newblock In: 2017 IEEE Wireless Communications and Networking Conference
  (WCNC), pp. 1--6 (2017)

\bibitem{qc_iot_whitepaper}
{Qualcomm Inc.}: How will 5g transform industrial iot.
\newblock Whitepaper available at
  https://www.qualcomm.com/media/documents/files/how-5g-will-transform-industrial-iot.pdf

\bibitem{raghavan2016}
Raghavan, V., Cezanne, J., Subramanian, S., Sampath, A., Koymen, O.:
  Beamforming tradeoffs for initial ue discovery in millimeter-wave mimo
  systems.
\newblock IEEE J. Sel. Topics Signal Process. \textbf{10}(3), 543--559 (2016)

\bibitem{Rappaport:02}
Rappaport, T.S.: Wireless Communications: Principles and Practice, second edn.
\newblock Prentice Hall, Upper Saddle River, NJ (2002)

\bibitem{rave_game2019}
{Rave}, W., {Khalili-Marandi}, M.: The elimination game or: Beam selection
  based on m-ary sequential competition elimination.
\newblock In: WSA 2019; 23rd International ITG Workshop on Smart Antennas
  (2019)

\bibitem{Samimi:AoAD}
Samimi, M., Wang, K., Azar, Y., Wong, G.N., Mayzus, R., Zhao, H., Schulz, J.K.,
  Sun, S., Gutierrez, F., Rappaport, T.S.: 28 {GHz} angle of arrival and angle
  of departure analysis for outdoor cellular communications using steerable
  beam antennas in {N}ew {Y}ork {C}ity.
\newblock In: Proc. IEEE VTC (2013)

\bibitem{sim_mlv2v2018}
{Sim}, G.H., {Klos}, S., {Asadi}, A., {Klein}, A., {Hollick}, M.: An online
  context-aware machine learning algorithm for 5g mmwave vehicular
  communications.
\newblock IEEE/ACM Transactions on Networking \textbf{26}(6), 2487--2500 (2018)

\bibitem{Singh2009Limits}
Singh, J., Dabeer, O., Madhow, U.: On the limits of communication with
  low-precision analog-to-digital conversion at the receiver.
\newblock IEEE Trans. Commun. \textbf{57}(12), 3629--3639 (2009)

\bibitem{Song2008}
Song, I., Jeon, J., Jhon, H., Kim, J., Park, B., Lee, J.D., Shin, H.: A simple
  figure of merit of {RF MOSFET} for low-noise amplifier design.
\newblock IEEE Electron Device Lett. \textbf{29}(12), 1380--1382 (2008)

\bibitem{cairecs2018}
Song, X., Haghighatshoar, S., Caire, G.: A scalable and statistically robust
  beam alignment technique for millimeter-wave systems.
\newblock IEEE Trans. Wireless Commun. \textbf{17}(7), 4792?4805 (2018)

\bibitem{souto_gen2019}
{Souto}, V.D.P., {Souza}, R.D., {Uchoa-Filho}, B.F., {Li}, Y.: A novel
  efficient initial access method for 5g millimeter wave communications using
  genetic algorithm.
\newblock IEEE Transactions on Vehicular Technology \textbf{68}(10), 9908--9919
  (2019)

\bibitem{VanTrees:01a}
{Van Trees}, H.L.: Detection, Estimation and Modulation Theory, Part I.
\newblock Wiley, New York, NY (2001)

\bibitem{wang2012}
Wang, Y., Afshar, B., Ye, L., Gaudet, V.C., Niknejad, A.M.: Design of a low
  power, inductorless wideband variable-gain amplifier for high-speed receiver
  systems.
\newblock IEEE Trans. Circuits and Syst. I: Reg Papers \textbf{59}(4), 696--707
  (2012)

\bibitem{xiu_outband2019}
{Xiu}, Y., {Wu}, J., {Xiu}, C., {Zhang}, Z.: Millimeter wave cell discovery
  based on out-of-band information and design of beamforming.
\newblock IEEE Access \textbf{7}, 23076--23088 (2019)

\bibitem{yan_cs2016}
{Yan}, H., {Cabria}, D.: Compressive sensing based initial beamforming training
  for massive mimo millimeter-wave systems.
\newblock In: 2016 IEEE Global Conference on Signal and Information Processing
  (GlobalSIP), pp. 620--624 (2016)

\bibitem{zhang_code2017}
{Zhang}, J., {Huang}, Y., {Shi}, Q., {Wang}, J., {Yang}, L.: Codebook design
  for beam alignment in millimeter wave communication systems.
\newblock IEEE Transactions on Communications \textbf{65}(11), 4980--4995
  (2017)

\bibitem{zhang_game2016}
{Zhang}, Q., {Saad}, W., {Bennis}, M., {Debbah}, M.: Quantum game theory for
  beam alignment in millimeter wave device-to-device communications.
\newblock In: 2016 IEEE Global Communications Conference (GLOBECOM), pp. 1--6
  (2016)

\bibitem{Rappaport:28NYCPenetrationLoss}
Zhao, H., Mayzus, R., Sun, S., Samimi, M., Schulz, J.K., Azar, Y., Wang, K.,
  Wong, G.N., Gutierrez, F., Rappaport, T.S.: 28 {GHz} millimeter wave cellular
  communication measurements for reflection and penetration loss in and around
  buildings in {N}ew {Y}ork {C}ity.
\newblock In: Proc. IEEE ICC (2013)

\end{thebibliography}

%
%

\end{document}